\documentclass[aps,onecolumn,10pt]{revtex4}
\usepackage{amsmath}
\usepackage{amssymb}
\usepackage{hyperref}
\hypersetup{colorlinks=true,linkcolor=red,citecolor=cyan,urlcolor=magenta}
\numberwithin{equation}{section}
\numberwithin{equation}{section}

\begin{document}
\allowdisplaybreaks
\setcounter{equation}{0}

\title{PT symmetry and the square well potential: Antilinear symmetry rather than Hermiticity in scattering processes}

\author{Philip D. Mannheim}
\affiliation{Department of Physics, University of Connecticut, Storrs, CT 06269, USA \\
philip.mannheim@uconn.edu\\ }

\date{May 4 2026}

\begin{abstract}
While a Hamiltonian with a real potential has bound states with real energy below the scattering threshold, it produces resonances with complex energies above threshold. Scattering states are not square integrable, being instead delta function normalized. This lack of square integrability breaks the connection between Hermiticity and real eigenvalues, to thus allow for real bound state sector eigenvalues and complex scattering sector eigenvalues. When written as contour integrals delta functions take support in the complex plane, with the scattering amplitude being able to take support in the complex plane too. However, the scattering amplitude is CPT symmetric (or PT symmetric if C is conserved), regardless of whether or not states are square integrable. For resonance scattering this antilinear symmetry requires the presence of a complex conjugate pair of energies, one to describe the excitation of the resonance and the other to describe its decay, with it being their interplay that enforces probability conservation. Each complex pair of energy eigenvalues corresponds to only one observable resonance not two, to thus modify the standard pure decaying complex energy pole discussion of resonances. We show that the non-relativistic real potential square-well Schr\"odinger equation  possesses PT symmetry in both the bound and scattering sectors, with there being complex conjugate pairs of energy eigenvalue solutions in the scattering sector. The Hamiltonian thus acts as as Hermitian operator below the scattering threshold and as a non-Hermitian one above it. For those values of the potential for which bound states lie right at the top of the well the scattering amplitude threshold branch point is an exceptional point, a characteristic of systems with antilinear symmetry at which there are more independent solutions to the Schr\"odinger equation than there are eigenstates of a then non-diagonalizable and thus again non-Hermitian Hamiltonian. The square well  thus provides an explicit realization of how antilinearity is more general than Hermiticity. 
\end{abstract}

\maketitle

\section{Overview and   Objectives}
 \label{S1}

 In this introduction we present an overview of the objectives of this paper and summarize our findings. To begin, we recall the standard treatment of the three-dimensional, non-relativistic, spherically symmetric  square-well Schr\"odinger problem for a particle of mass $m$ in a finite-depth potential $V(r<a)=0$, $V(r>a)=V_0$ with radial coordinate $r$ and real, positive and finite $V_0$. In the complex energy plane the scattering amplitude $f(E)$ for scattering off the square well has poles on the real axis with $E<V_0$,  has a branch point at the scattering threshold $E=V_0$, has a branch cut that extends from the branch point  to infinite $E$, and has poles at $E=E_0-i\Gamma$ with $E_0>V_0$ and $\Gamma>0$.  The below-threshold real $E$ axis states are stationary eigenstates with spatially bounded eigenfunctions. Above threshold where ${\rm Re}[E]>V_0$ there are two kinds of states,  a continuum of spatially unbounded stationary states with real $E$ that lie all along the branch cut, together with states with $E=E_0-i\Gamma$ where $E_0>V_0$  that are not stationary eigenstates.  This structure  raises two concerns. The first is how a Hamiltonian such as that associated with a square well with real $V_0$ could have non-stationary eigenstates with complex eigenvalues, with the second being what then happens to probability conservation when the Hamiltonian does have them. In the present paper we shall address both of these issues.
 
In attempting to do so we have studied and analytically solved the square-well potential Schr\"odinger equation, and without any reference to the scattering amplitude  we have found two additional, apparently  novel features that its solutions possess.  Firstly, we have not only found exponentially decaying non-stationary energy eigenstate solutions with  energy $E=E_0- i\Gamma$, each one of them is accompanied by another equally non-stationary energy eigenstate solution with energy $E=E_0+i\Gamma$, viz. one that is exponentially growing in time rather than decaying. These states are bona fide energy eigenstates since $i\partial_t e^{-i(E_0\pm i\Gamma) t}=(E_0\pm i\Gamma)e^{-i(E_0\pm i\Gamma) t}$. As described in Sec. \ref{S5a}, to establish this result we have analytically continued the energy dependence of the standard square-well bound state equation that fixes the bound state energies (just as one analytically continues the scattering amplitude in energy). And since all the coefficients in the bound state equation are real, on  continuing it into the scattering region we can only find complex conjugate pairs of energy eigenstate solutions to the Schr\"odinger equation, with it not being possible for there to be just the standard $E_0-i\Gamma$ type solutions alone. With these complex energy eigenvalues the Hamiltonian with its real potential  is not Hermitian.

In addition, we find that  all the bound state and complex conjugate energy pair scattering state  eigenfunctions have $\tan\delta=-i$, where $\delta$ is  the energy-dependent square-well scattering sector region wave function phase shift. Since the scattering amplitude is of the form $f(E)=\tan\delta/(1-i\tan\delta)$ we see that all of these particular Schr\"odinger equation energy eigenvalues are poles in the scattering amplitude. Thus the standard analytic pole  term structure of the scattering amplitude in the complex energy plane originates in the  structure of solutions to  the Schr\"odinger equation. Also there are plane wave continuum states with real energy, and they too are solutions to the Schr\"odinger equation. However, they have real $\delta$ (i.e., not $\tan\delta=-i$) and real momenta, and in Sec. \ref{S4} we will show that they associated with the cut (i.e., not the pole structure) of $f(E)$.

However, since we do solve the Schr\"odinger equation directly, in (\ref{5.10a}) we also determine the radial dependence of the eigenfunctions, something that cannot be inferred from the structure of the scattering amplitude alone since it only depends on the energy. Since the momenta of the wave functions associated with the complex $E_0\pm i\Gamma$ energy eigenvalues also have to be complex ($e^{i(k_1\pm ik_2)r}$), the radial dependence is not just oscillating but also has terms that grow or fall exponentially in the radial coordinate. The modes that decay exponentially  in time are accompanied by a radial behavior that  grows exponentially (so-called Gamow vectors), while the modes that grow exponentially  in time are accompanied by a radial behavior that  falls exponentially (which we label anti-Gamow vectors).  The standard treatment of Gamow vectors is to place them in a dual space $\Phi^*$ of a Gelfand triple rigged Hilbert space $\Phi \subset{\mathcal H}\subset \Phi^*$ with a sufficiently convergent set of test functions in the test function space $\Phi$ so that the  interplay of $\Phi^*$ and $\Phi$ in a functional generalization of the standard Dirac bra ket yields normalizability.  However, since we have both a Gamow vector and an anti-Gamow vector we do not need to do this. Rather, we can use the techniques of the $PT$-symmetry program where the Hamiltonian commutes with the $PT$ operator and where there is then a $\hat{V}$ operator that effects the pseudo-Hermiticity condition $\hat{V}\hat{H}\hat{V}^{-1}=\hat{H}^{\dagger}$  with each $\hat{H}$ dynamically determining its own $\hat{V}$. Then as we show in (\ref{5.13a}) and Sec. \ref{S6} one can construct a $\hat{V}$-based probability amplitude for the square well that is  time independent, viz. $\langle \psi_1\vert \hat{V}\vert \psi_2\rangle$ rather than the standard pre-assigned $\langle \psi_1\vert \psi_2\rangle$. The form of $\langle \psi_1\vert \hat{V}\vert \psi_2\rangle$ for the square well is such that  the exponentially growing and exponentially falling time behaviors cancel each other, while at the same time  the exponentially growing and exponentially falling radial behaviors not only also cancel each other, they lead to the same  asymptotic radial behavior as that of plane waves with real momenta.  (With just the $E_0-i\Gamma$ mode with its exponentially falling time behavior,  its exponentially growing radial  behavior would not be acceptable.) The existence of complex conjugate pairs of energy eigenvalues is thus fully viable physically.   As we see, in the $PT$-symmetry  framework the anti-Gamow vector serves as the test function without any need to rig the Hilbert space.

In regard to the second novel feature that we have found,  we show in Sec. \ref{S5b} that on taking the $\Gamma $ goes to zero limit we find that the pair of non-stationary eigenstates with energies $E_0\pm i\Gamma$ collapses onto just one  common single eigenstate with $\Gamma=0$ and $E_0=V_0$. This occurs when bound states  lie right at the top of the well, i.e., lie right  at the scattering amplitude threshold $E=V_0$, something that can occur only for  values of the potential that obey $V_0=\pi^2\hbar^2/8ma^2,~9\pi^2\hbar^2/8ma^2,~25\pi^2\hbar^2/8ma^2,\ldots$ Thus  the Hamiltonian loses an eigenstate at  this point, with the threshold branch point  being an exceptional point at which the eigenspectrum of the Hamiltonian becomes incomplete,  with there  being more independent solutions to the Schr\"odinger equation than there are eigenstates of the then non-diagonalizable and thus non-Hermitian Hamiltonian. At the exceptional point the eigenstate that has been lost becomes  a non-stationary state that grows linearly in time. Specifically, in the $\Gamma \rightarrow 0$ limit we have 
\begin{align}
\frac{(e^{-i(E_0-i\Gamma) t}+e^{-i(E_0+i\Gamma) t})}{2}\rightarrow e^{-iE_{0} t}, \qquad \frac{(e^{-i(E_0-i\Gamma) t}-e^{-i(E_0+i\Gamma) t})}{-2\Gamma} \rightarrow te^{-iE_{0} t}.
\label{1.1}
\end{align}
(Even while the $E_0-i\Gamma$ and $E_0+i\Gamma$ modes are both separately eigenstates of the Hamiltonian, linear combinations of them are not, with only their sum collapsing onto an eigenstate in the $\Gamma \rightarrow 0$ limit, but not their difference as divided by $-2\Gamma$.) The modes that grow linearly in time are foreign to the scattering amplitude framework, are not associated with it, and cannot be found from it by Fourier transform. The two novel  features that we present here for  a system  as familiar  as  the square well do not appear to have previously been reported in the literature. 

The square well thus loses eigenstates at the exceptional point and has non-real eigenvalues in the scattering region. In both of these two sectors  the connection between Hermiticity and real eigenvalues is lost even though $V_0$ remains real throughout. While the square-well Hamiltonian is self-adjoint when acting on the square-integrable bound state eigenfunctions, in the scattering region the Hamiltonian is not  self-adjoint since one cannot throw surface terms away in an integration by parts as the scattering region eigenfunctions are not  square integrable, especially ones that grow exponentially in space. [For the momentum $p=-i\hbar\partial_x$ for instance we have $(-i\hbar \int dx \psi_1^*\partial_x\psi_2)^*+i\hbar \int dx \psi_2^*\partial_x\psi_1=i\hbar(\psi^*_2\psi_1)|^{\infty}_{-\infty}\neq 0$.)] At the exceptional point completeness of the eigenbasis is lost, and accordingly so is Hermiticity. 

In the scattering sector  wave functions $\psi(x)$ obey the relation  (as derived below as (\ref{3.12}))
\begin{align}
\langle \psi\vert\hat{H}\vert \psi\rangle-\langle \psi\vert\hat{H}^{\dagger}\vert \psi\rangle=(E-E^*)\langle \psi\vert\psi\rangle=(E-E^*)\int_{-\infty}^{\infty} dx \langle \psi\vert x\rangle\langle x\vert\psi\rangle=(E-E^*)\int_{-\infty}^{\infty} dx\psi^*(x)\psi(x),
\label{1.2}
\end{align}
with all the matrix elements in (\ref{1.2}) being infinite since scattering sector  wave functions are unbounded. Thus  one cannot use (\ref{1.2})  to establish the reality of eigenvalues. Thus in the scattering sector the standard connection between Hermiticity and real eigenvalues is lost and complex energy eigenvalues are not forbidden. While not forbidden the eigenvalues could still be real. However, the delta functions used to  normalize plane waves  are not ordinary functions. Rather, they can be treated as distributions, or they and the theta functions that are related to them can be represented as  complex plane contour integrals. Rather than treat them as distributions in a rigged Hilbert space, in Sec. \ref{S4} we shall instead treat them as  contour integrals. It is then in this way that we can obtain complex eigenvalues in the scattering sector.

The general structure of the solutions that we have found can be understood from various square well limits. For the free theory (viz. $V_0=0$) there is only a continuum of  plane waves and the Hamiltonian is not self-adjoint. For the infinitely deep well (viz. $V_0=\infty$) wave functions are only non-zero inside the well, and matching boundary conditions at the edge of the well produce bound states, with the Hamiltonian being self-adjoint. For the finite-sized well there are bound states inside the well where the Hamiltonian is self-adjoint, and in the  scattering sector there again is a plane wave continuum, though a now phase-shifted one  with  real phase shift, and  the Hamiltonian is again not self-adjoint. However, in the same way that matching boundary conditions at the edge of the well produce bound states in the well, above threshold they produce complex conjugate pairs of energy eigenstates, with the Hamiltonian not being Hermitian.   In addition, at exceptional points there are solutions to the Schr\"odinger equation that satisfy the matching boundary conditions but are  not eigenstates of the Hamiltonian at all. 

While we have noted that the free theory Hamiltonian is not self-adjoint when it acts on plane waves, it can be made so in a rigged Hilbert space, one that is constructed to handle the distribution properties of delta function normalized plane waves. Hermiticity is then secured and all energy eigenvalues are then real. For the infinitely deep square well all energy eigenvalues are also real and the Hamiltonian is again Hermitian. However, things change for the finite-depth square well, a system that  has both a bound state sector and a scattering sector. While the bound state energy eigenvalues are still real, in the scattering sector as well as plane wave  eigenfunctions  with real energy, there are also energy eigenstates in complex conjugate pairs, something not achievable with a Hermitian Hamiltonian. The associated radial wave functions are not just oscillating but grow or fall exponentially. And even if, just like the continuum modes, the radially growing Gamow vectors are placed in a Hilbert space that is rigged with an appropriately convergent set of test functions, that would only affect the behavior of the radial wave function normalization and would not make the energy eigenvalues real.   Thus in the complex conjugate pair sector the Hamiltonian is not just not self-adjoint if one uses the $\langle \psi_1\vert\psi_2\rangle$ inner product, with exponential radial growth it is very far from being so. This lack of self-adjointness goes hand in hand with the existence of non-real energy eigenvalues. Also we should note that the mode that grows exponentially in time actually falls exponentially in space. It thus is  $\langle \psi_1\vert\psi_2\rangle$  square-integrable. And thus even if one were to restrict the solutions to those that are square integrable, one would still have to include the exponentially growing in time one. Physically this would be unacceptable. Thus it is only by including both the $E_0+i\Gamma$ and $E_0-i\Gamma$ modes and recognizing that the Hamiltonian while not Hermitian is nonetheless pseudo-Hermitian that one can construct a $\langle \psi_1\vert \hat{V} \vert \psi_2\rangle$  probability amplitude that neither grows in time or in space. For the square well Hamiltonian we see that Hermiticity depends on energy. Below threshold the Hamiltonian is Hermitian, but above threshold it is not, being instead pseudo-Hermitian.

There is an equivalent though more unifying way to describe the situation. As we discuss in \cite{Mannheim2018a} and Sec. \ref{S6} below,  for any Hamiltonian one should in general use as inner product the $\langle L \vert R \rangle$ overlap of its right- and left-eigenvectors $\vert R\rangle$ and $\langle L \vert$. If the Hamiltonian is Hermitian then $\langle L \vert=\langle R \vert$, while if the Hamiltonian is pseudo-Hermitian then $\langle L \vert=\langle R \vert \hat{V}$. This is a unifying description since it applies for both real and complex conjugate energy eigenvalues and to both bound and scattering sectors, with the Hamiltonian determining the appropriate inner product in each case. Thus the general approach of the $PT$-symmetry program is to use as inner product $\langle L \vert R \rangle$, a more general and not pre-assignable choice (each Hamiltonian determines its own $\langle L \vert$ and $\vert R \rangle$) than the standard pre-assigned $\langle R \vert R \rangle$.

The eigenspectrum pattern that we obtain for the square well (real energies in the bound state sector, an exceptional point at the scattering threshold,  and complex conjugate pairs of energy eigenvalues in the scattering sector) is actually a familiar one, as it can occur when a Hamiltonian has an antilinear symmetry. Thus while self-adjointness is lost in the scattering sector,  antilinearity is not sensitive to self-adjointness and applies to all sectors and is thus more general than Hermitcity.  In Sec. \ref{S8} we present a simple two-dimensional $PT$-symmetric matrix $M(s)$. As we change the parameter $s$ we uncover the above pattern of energy eigenvalues. Regardless of square-integrability issues this same pattern must repeat in the infinite-dimensional $PT$-symmetric case with it being the energy that is then the variable parameter. And indeed for the square well we find this very pattern, to thus confirm that in the scattering sector there must be complex conjugate pairs of energy eigenvalue solutions to the Schr\"odinger problem, with the Hamiltonian not being Hermitian in this sector.

In general, we note that if a Hamiltonian $\hat{H}$ has a real potential, the determinantal condition $\vert  \hat{H}-I \lambda \vert=0$  that fixes energy eigenvalues is a real equation. Consequently, eigenvalues can then either be real or occur in complex conjugate pairs. If the eigenspectrum of $\hat{H}$ is complete and $\hat{H}$ is Hermitian then all eigenvalues are real, while if instead $\hat{H}$ is pseudo-Hermitian, viz. $\hat{V}\hat{H}\hat{V}^{-1}=\hat{H}^{\dagger}$, then $\hat{H}$ and $\hat{H}^{\dagger}$ have the same set of eigenvalues, and eigenvalues can then only be real or in complex conjugate pairs. Thus while we cannot associate a Hermitian Hamiltonian with complex conjugate eigenvalue pairs, we can associate a pseudo-Hermitian one with them. Finally, if the eigenspectrum of $\hat{H}$ is incomplete (which it can be since the condition $\vert \hat{H}-I\lambda  \vert=0$ contains less information than the Schr\"odinger equation $\hat{H}\vert \psi\rangle=E\vert\psi \rangle$ itself), then the Hamiltonian is of non-Hermitian Jordan-block form. Taken together all of these possible options precisely fit  the pattern we find for the real square well potential. As we see, without first determining  the eigenspectrum one cannot guarantee that a Hamiltonian is Hermitian even if it has a real potential. Moreover, given an arbitrary Hamiltonian it is not always possible to explicitly determine the eigenspectrum or  construct an explicit $\hat{V}$ that effects the pseudo-Hermitian condition $\hat{V} \hat{H}\hat{V}^{-1}= \hat{H}^{\dagger}$. However, without needing to do any of this, one can readily check if a Hamiltonian has an antilinear symmetry such as $PT$ since pseudo-Hermiticity then follows \cite{Mannheim2018a}, with this being the case for the square well.

As such, our current square-well study provides an explicit realization of  our contention   in \cite{Mannheim2018a} that it is antilinearity rather than Hermiticity that should be taken as the guideline for quantum theory.  In support of this proposal we note  that in \cite{Mannheim2018a} we had predicted that theories that have a time delay $\hbar/\Gamma$ (viz. one associated with $E_0-i\Gamma$) should  also have an equal and opposite  time advance (viz. a negative time delay $-\hbar/\Gamma$ associated with $E_0+i\Gamma$). As  discussed in \cite{Mannheim2025} and below in Sec. \ref{S7}, such a time advance has  recently been detected  in atomic scattering experiments \cite{Sinclair2022,Angulo2024}, to thus lend support to the centrality of antilinearity in quantum theory that we had proposed. Due to the fact that  the $CPT$ theorem as originally established for Hermitian systems has been shown \cite{Mannheim2018a} to apply to non-Hermitian systems as well ($C$ is charge conjugation, $P$ is parity and $T$ is antilinear time reversal), it follows that whenever there is a decaying state with energy $E_0-i\Gamma$, because of the antilinear nature of $CPT$ symmetry there must be an accompanying growing state with energy $E_0+i\Gamma$. The former of these two states gives rise to a time delay and the latter one gives rise to an equal and opposite  time advance. As such, the existence of complex conjugate pairs of energy eigenvalues dates back to Wigner's study \cite{Wigner1960} of time reversal invariance, a symmetry that just like $CPT$ itself is antilinear.  In \cite{Mannheim2018a} and  \cite{Simon2019} it was shown that this aspect of $PT$ symmetry translates into complex energy plane poles in the scattering amplitude.  

It is the existence of both of the $E_-=E_0-i\Gamma$ and $E_+=E_0+i\Gamma$ energy eigenvalues that enforces  probability conservation. Specifically, if we define $\psi_{\pm}(t)=e^{-i(E_0\pm i\Gamma)t}$, we find that both $\psi_{+}^*(t)\psi_+(t)=e^{2\Gamma t}$ and  $\psi_{-}^*(t)\psi_-(t)=e^{-2\Gamma t}$ are time dependent, but $\psi_{+}^*(t)\psi_-(t)$ and  $\psi_{-}^*(t)\psi_+(t)$ are time independent. Thus in general, and  in some specific examples that we give below, examples in which the coefficients of the time-dependent combinations are zero, we shall see that  the decaying and growing modes regulate each other so as to maintain probability conservation. Thus as the number of  levels in a decaying system is depleted, the number of levels in the decay products of that system grows so that the total number of levels remains constant. From the perspective of  the scattering amplitude  the two poles at  $E_-=E_0-i\Gamma$ and $E_+=E_0+i\Gamma$ can only regulate each other this way if they are in the same contour in the  complex energy plane, a contour that thus cannot run along the real $E$ axis  but would have to be deformed so as to enclose both poles.  Even though there are then two poles on the same Riemann sheet, as had been noted in \cite{Mannheim2018a} and as will be seen in Sec. \ref{S6a} below, this only gives rise  to one observable resonance and not two. In regard to the  need to deform the contour,   in Sec. \ref{S6a} we also show that putting the two complex conjugate poles in the same contour (i.e., on the  same Riemann sheet) is necessary for causality. (While solving the Schr\"odinger equation gives complex conjugate pairs of energy eigenvalues, in and of itself that does not tell us on which Riemann sheet in  the scattering amplitude to locate them.) 

The fact that there is only one observable resonance is of significance for a different reason. In the standard analysis of the square well one studies the scattering cross-section as a function of real energy (viz. the experimental situation) and finds resonances. At these resonances the phase shift obeys $\tan\delta =\pi/2$, and the scattering amplitude can be approximated by the Breit-Wigner forms for $f(E)$ and  $D_{\rm BW}(E)$ given in (\ref{2.1}) and (\ref{2.2}) below, viz.
\begin{align}
f_{\rm BW}(E)= \frac{\Gamma}{E_0-i\Gamma-E}=\frac{\Gamma(E_0-E)+i\Gamma^2}{(E_0-E)^2+\Gamma^2},\qquad 
D_{\rm BW}(E)=\frac{1}{E-E_0+i\Gamma}
=\frac{E-E_0-i\Gamma}{(E-E_0)^2+\Gamma^2},
\label{1.3}
\end{align}
as evaluated with real $E$. Such a form then implies that the scattering amplitude has a complex pole at $E=E_0-i\Gamma$, assuming that is that we can make an  analytic continuation from real $E$ to the $E=E_0-i\Gamma$ pole without encountering any other contribution or impediment on the way. However, the continuation  of the Breit-Wigner form need not coincide with the continuation of the scattering amplitude itself. (For instance, apart from, say, possible cuts,  for real $E$ there could be terms of a form such as $(E_0-E)/\Gamma$ in $\tan\delta$ that could be negligible near $E=E_0$ but important in the complex $E$ plane.) 

Now a scattering amplitude can be expressed in terms of the resolvent operator $1/(E-\hat{H})$, a quantity that correlates poles with eigenvalues for  real or complex $E$. However, any such correlation can only obtained for  matrix elements of the resolvent that are evaluated in  energy eigenstates of the Hamiltonian. Thus the standard understanding of the Breit-Wigner $f_{\rm BW}(E)$ is that there are in fact no energy eigenstates associated with the $E_0-i\Gamma$ poles. In consequence, the presence of the Breit-Wigner $E_0-i\Gamma$ pole in the scattering amplitude had not been thought to bear on the Hermiticity of the Hamiltonian (in a rigged Hilbert space as needed for the continuum modes), with a possible loss of Hermiticity not being contemplated. Rather, the presence of a complex pole was thought to be solely a property of the analytic structure of the scattering amplitude in the complex energy plane, given the fitting to the resonance shape with a Breit-Wigner form as evaluated with real energy, with the Hermiticity of the square well Hamiltonian not being questioned. 

However, since in our current paper we find  a complex conjugate pair of energy eigenvalue solutions to the Schr\"odinger equation, we establish that Hermiticity is in fact lost. Nonetheless, despite the presence of complex conjugate pairs of energy eigenvalues we not only find that there is only one resonance per pair, the shape of the scattering amplitude as given in the $PT$-symmetry-derived $D_{ PT}(E)$ in (\ref{6.7}) below, viz.
\begin{align}
D_{PT}(E)=\frac{1}{E-(E_0-i\Gamma)}-\frac{1}{E-(E_0+i\Gamma)}=\frac{-2i\Gamma}{(E-E_0)^2+\Gamma^2},
\label{1.4}
\end{align}
is identical in form to the Breit-Wigner profile at the resonance energy. If we thus identify the resonance structure not with $D_{\rm BW}(E)$ but with $D_{ PT}(E)$, we can now directly associate its structure with the complex conjugate pairs of poles in the scattering amplitude that are due to the complex energy eigenvalues of the  Schr\"odinger equation.  To make contact with resonances in the real $E$ scattering amplitude we introduce phase shifts $\delta_-(E)$ and $\delta_+(E)$ of the form
\begin{align}
\tan\delta_-(E)=\frac{\Gamma}{E_0-E},\qquad \tan\delta_+(E)=-\frac{\Gamma}{E_0-E}.
\label{1.5}
\end{align}
At the poles they obey $\tan\delta_-(E_0-i\Gamma)=-i$, $\tan\delta_+(E_0+i\Gamma)=-i$, just as poles should. At real $E=E_0$ they obey $\delta_-(E_0)=\pi/2$, $\delta_+(E_0)=-\pi/2$, just as resonances should, while respectively giving the time delay and time advance described above. Together, they give a scattering amplitude
\begin{align}
f_{PT}(E)=\frac{\tan\delta_-(E)}{1-i\tan\delta_-(E)}+\frac{\tan\delta_+(E)}{1-i\tan\delta_+(E)}=\frac{\Gamma}{E_0-i\Gamma-E}-\frac{\Gamma}{E_0+i\Gamma-E}=\frac{2i\Gamma^2}{(E-E_0)^2+\Gamma^2}.
\label{1.6}
\end{align}
We recognize (\ref{1.6}) as being of the same form as $D_{PT}(E)$.  Also we note that since $\delta_+(E)=-\delta_-(E)$, it follows that $f_{PT}(E)=e^{i\delta_-(E)}\sin\delta_-(E)-e^{-i\delta_-(E)}\sin\delta_-(E)=2i\sin^2\delta_-(E)$, to thus give a real energy resonance at $\delta_-(E)=\pi/2$, viz. $E=E_0$. The presence of resonances in the real energy scattering cross-section thus indicates that the Hamiltonian is not Hermitian, and that the continuation not of $D_{\rm BW}(E)$ but of $D_{ PT}(E)$ between real $E$ and the complex poles is both needed and valid. 

The contrast here with the standard Breit-Wigner discussion is as follows. For the Breit-Wigner case one starts with a real energy resonance and continues $f_{\rm BW}(E)$ into the complex energy plane, with there being no complex energy eigenstate associated with the Breit-Wigner pole. In the $PT$ case described here one instead starts with the Schr\"odinger equation, obtains bona fide complex conjugate pairs of energy eigenstates, and then continues the ensuing $f_{PT}(E)$ scattering amplitude from the associated complex energy plane poles to real $E$ and obtains bona fide resonances. This is significant since high energy scattering cross section resonances are identified with actual physical particles, viz. actual physical  eigenstates of the scattering Hamiltonian and actual physical poles of the resolvent operator $1/(E-\hat{H})$. Thus they cannot be associated with a purely mathematical continuation of the scattering amplitude. Rather, a posteriori, we see that they should be associated with not one complex pole but with a complex conjugate pair of poles just as antilinear symmetry requires. Also, as we shall show in Sec. \ref{S5c}, despite the common notation used for $f_{\rm BW}(E)$ and $f_{PT}(E)$, the $E_0$ and $\Gamma$ determined from $f_{\rm BW}(E)$ and $f_{PT}(E)$ are different, and do not continue into each other or bear any relation to each other. (In the numerical example referenced in Sec. \ref{S5c} the value determined for the Breit-Wigner $\Gamma$ is actually negative not positive.) Since $f_{\rm BW}(E)$ and $f_{PT}(E)$  do not continue into each other, fits to scattering cross sections can only be associated with physical particles  if they are made using $f_{PT}(E)$. However, since $f_{\rm BW}(E)$ and $f_{PT}(E)$ agree on resonance, fits using either would experimentally lead to the same $E_0$, though they could give different determinations for $\Gamma$ dependent on the degree to which  they may or may not  differ near resonance.

That the  $E_-=E_0-i\Gamma$ and $E_+=E_0+i\Gamma$ modes do regulate each other so as to give probability conservation is actually a consequence of the $CPT$ theorem. Specifically, in \cite{Mannheim2018a} the $CPT$ theorem was derived under just two conditions: invariance under the complex Lorentz group and probability conservation. With $CPT$ being an antilinear operator, complex energy modes must exist in complex conjugate pairs, pairs  that will necessarily regulate each other so as to give probability conservation. In regard to the $CPT$ theorem we note also that in the charge-conjugation-conserving, non-relativistic limit $CPT$ defaults to $PT$. Since in a non-relativistic experiment the observer is nonetheless free to move with a relativistic velocity, any $C$-conserving non-relativistic experiment that can be performed in a laboratory must still be compatible with relativity, and hence must be $PT$ symmetric. Thus in the following we shall restrict to antilinear $PT$ symmetry.  However, even with $PT$ symmetry one could have both $P$ and $T$ conserved, or both $P$ and $T$ not conserved. The square well of interest to us in the paper falls into the both $P$ and $T$ conserved category. 

While $C$-conserving situations in which one of $P$ or $T$ is conserved and the other not would violate the  $CPT$ theorem, study of such systems can still be instructive. And in \cite{Simon2019} various combinations of conserved or not conserved $P$, $T$ or $PT$ symmetries have been studied quite comprehensively, and the regions in the complex energy plane where there are to be scattering amplitude singularities were identified. In general though we should note that knowing only that are such regions does not in and of itself determine the explicit locations of the singularities within those regions, or even guarantee that such singularities even exist.  Absent some further information  it only determines in which quadrants in the complex energy plane singularities will be found if they do exist. For instance, for a generic complex conjugate pair of scattering amplitude poles of the form $E_0\pm i\Gamma$ one can only say that the $E_0+i\Gamma$ pole lies in the ${\rm Re}[E]>0$, ${\rm Im}[E]>0$ quadrant in the complex energy plane, while the $E_0-i\Gamma$ pole lies in the ${\rm Re}[E]>0$, ${\rm Im}[E]<0$ quadrant.  Moreover, just as with bound states, to determine specific values for $E_0$ and $\Gamma$ requires a study of the Schr\"odinger equation associated with the potential that is producing the scattering in the first place. Moreover, one should not expect to find singularities for arbitrary values of the parameters in  the potential even if one knows in which complex energy plane quadrant they would have to appear when they do exist. Thus for the three-dimensional, spherically symmetric square well of interest to us here there are no bound states at all if $V_0<\pi^2\hbar^2/8ma^2$. And as we show in Sec. \ref{S5a}, for there to be a complex conjugate pair of poles at all, non-trivial solutions to (\ref{5.7a}) and (\ref{5.8a}) below require that $(2m/\hbar^2)^{1/2}a\mu>\pi$ where $E_0\pm i\Gamma=(\mu\pm i\nu)^2$. For the finite-depth attractive square well  it is generally thought that there will be resonances in the real energy scattering cross section. However, without studying the Schr\"odinger equation as done in this paper one a priori cannot say where they would be or even whether for specific values of parameters there would be any at all. Thus without studying  the Schr\"odinger equation one does not know if there are to be resonances in the scattering cross section at all. The issue being raised here is that one typically uses scattering experiments to see whether and where there are resonances, and then uses that information to learn about the nature of the scattering potential. Since our work starts with the potential it shows what sorts of scattering amplitudes can occur, and finds them to be richer than commonly thought, with one having to fit resonance structures in cross sections using  $D_{ PT}(E)$ rather than $D_{\rm BW}(E)$.
 
While there can be complex energies in the scattering sector, in and of itself that is not sufficient to ensure probability conservation as states with energies of the form $E_0-i\Gamma$ decay and take  probability with them. If such systems are open systems the inclusion of the larger system in which they would then be contained could restore probability conservation (or possibly even Hermiticity).  However, with antilinear symmetry energies of the form $E_0+i\Gamma$ are contained within the system that contains the $E_0-i\Gamma$ modes. Systems that contain both the $E_0-i\Gamma$ and $E_0+i\Gamma$ modes are not open but are  self-contained closed systems. As had been noted above, the loss of population of a level as it decays ($E_0-i\Gamma$) leads to growth ($E_0+i\Gamma$) in the population of the level that it decays into. Probability conservation then requires that the total population of the two levels combined remains constant. Hence the discussion we provide of decays differs from the standard discussion as the standard discussion  is presupposed to occur in an open system context, while in the $PT$ case decays occur in a closed system context that also includes the growing modes. Below we will see that elastic scattering processes can also be treated as closed systems. 

Our paper is organized as follows. In Sec. \ref{S2} we briefly describe the standard single complex pole Breit-Wigner description of resonances. In Sec. \ref{S3} we discuss some general properties of plane waves. In Sec. \ref{S4} we show the $\delta(x-x^{\prime})$ normalization of position operator eigenstates has a  complex momentum plane structure, a structure  that is reflected in the branch cut structure of the scattering amplitude. In Sec. \ref{S5} we present the solution we have found to the square-well Schr\"odinger equation, showing  that in the scattering region there are Schr\"odinger equation eigenstates that appear in complex conjugate pairs. Also we explicitly construct a solution that as in (\ref{1.1})  grows linearly time, doing so though only for those values of the potential that are given by  $V_0=\pi^2\hbar^2/8ma^2,~9\pi^2\hbar^2/8ma^2,~25\pi^2\hbar^2/8ma^2,\ldots$ In Secs. \ref{S5} and \ref{S6}   we  provide a brief description of the $PT$ symmetry and pseudo-Hermiticity programs. In Sec. \ref{S6} we show how one can have two complex conjugate poles in the scattering amplitude and yet only have one observable resonance. Also in Sec. \ref{S6} we show how we can have probability conservation in the presence of non-stationary states.  In Sec. \ref{S7} we provide two applications of our ideas: spontaneous atomic emission and elastic scattering. In Sec. \ref{S8} we show how our study enlarges the scope and reach  of $PT$ symmetry.  In Sec. \ref{S9} we present some final comments. However, in order to appreciate the ideas we present in this paper we begin with the standard discussion of decays \cite{footnote0}.

\section{The Standard Breit-Wigner Approach}
\label{S2}

In the conventional treatment of the  quantum mechanics of a non-relativistic, real potential well one has bound states in the well and scattering states above it.  Each resonance that occurs in a scattering above the well is associated with an energy-dependent phase shift of the form $\tan \delta=\Gamma/(E_0-E)$, with $\delta$ being equal to $\pi/2$ when the energy $E$ is at resonance $E=E_0$. With a phase shift of this form the scattering amplitude behaves as 
\begin{align}
f(E)\sim e^{i\delta}\sin\delta= \frac{\Gamma}{E_0-i\Gamma-E},
\label{2.1}
\end{align}
and the propagator has the standard Breit-Wigner form
\begin{align}
D_{\rm BW}(E)=\frac{1}{E-E_0+i\Gamma}
=\frac{E-E_0-i\Gamma}{(E-E_0)^2+\Gamma^2},
\label{2.2}
\end{align}
with a pole at $E=E_0-i\Gamma$. To determine the associated time dependence we Fourier transform and obtain
\begin{align}
D_{\rm BW}(t)=\frac{1}{2\pi}\int _{-\infty}^{\infty}dE e^{-iEt}D_{\rm BW}(E)=\frac{1}{2\pi }\int _{-\infty}^{\infty}dE \frac{e^{-iEt}}{E-E_0+i\Gamma}.
\label{2.3}
\end{align}
With the pole being in the lower half of the complex $E$ plane, and with $e^{-iEt}$ vanishing on the lower-half plane circle at infinity when $t>0$ while vanishing on the upper-half plane circle at infinity when $t<0$, contour integration gives the causal, forward in time propagator 
\begin{align}
D_{\rm BW}(t)=\frac{1}{2\pi }\int _{-\infty}^{\infty}dE \frac{e^{-iEt}}{E-E_0+i\Gamma}=-i\theta(t)e^{-iE_0t-\Gamma t}.
\label{2.4}
\end{align}
Thus  $f(t)$ and $D_{\rm BW}(t)$ only possess a decaying mode, with a time delay $\Delta T(E)=\hbar d\delta/dE$ \cite{Wigner1955} off resonance and on resonance of the respective forms
\begin{align}
\Delta T(E)=\hbar\frac{d\delta}{dE}=\frac{\hbar\Gamma}{[(E-E_0)^2+\Gamma^2]},\qquad \Delta T(E_0)=\frac{\hbar}{\Gamma}.
\label{2.5}
\end{align}
This time delay is due to the fact that the wave is held by the well for a time $\Delta T(E)$.

With the potential being taken to be Hermitian the bound states have real energies. Despite this, scattering can produce resonances with complex energies and a non-stationarity associated with  time delays, to thus pose the two questions of how a Hermitian potential could produce non-stationary resonance states at all, and what happens to probability conservation when it does. To address these two issues we examine the status of plane waves.

\section{Status of Plane Waves}
\label{S3}

In an elastic scattering experiment by a localized source the asymptotic incoming states are plane waves and the asymptotic outgoing states are phase-shifted plane waves, with the phase shift being caused by the interaction of the incoming plane waves with the localized source. However, unlike bound states, which have square-integrable wave functions, plane waves are not only not square integrable they are normalized to delta functions, to thus not belong to a Hilbert space composed of square-integrable wave functions. Despite their name, delta functions are not functions at all but can be treated as distributions in a rigged Hilbert space, or, for our purposes here,  can be represented by complex plane contour integrals.  It is this very nature of delta functions that builds a knowledge of the complex momentum plane into  plane waves and enables scattering to produce non-stationary states. 

While it is common to represent the momentum operator $\hat{p}$ as $-i\hbar\partial_x$ in the basis in which the position operator $\hat{x}$ is diagonal with eigenvalue $x$, one can only realize the $[\hat{x},\hat{p}]=i\hbar$ commutator this way when the commutator acts on square-integrable test functions, such as those that  describe bound states. This cannot be done for plane waves, even though it is common practice to do so in discussing scattering off say a square well potential, and even though $e^{ipx/\hbar}$ is an eigenstate of $-i\hbar\partial_x$ with eigenvalue $p$. 

Nonetheless, the standard discussion of the square well potential  can be recovered using the structure of the commutation algebra associated with the $[\hat{x},\hat{p}]=i\hbar$ commutator (see \cite{Messiah1964}).  We introduce position and momentum eigenstates and normalization and closure conditions of the form 
\begin{align}
\hat{x}\vert x\rangle =x\vert x\rangle, \quad  \hat{p}\vert p\rangle =p\vert p\rangle, \quad \langle x_1\vert x_2\rangle=\delta(x_1-x_2), \quad  \langle p_1\vert p_2\rangle=\delta(p_1-p_2),\quad \int dx \vert x\rangle \langle x\vert =I, \quad \int dp \vert p\rangle \langle p\vert =I.
\label{3.1}
\end{align}
We introduce the translation operator $\hat{S}(a)=e^{-ia\hat{p}/\hbar}$ where $a$ is a constant, and obtain
\begin{align}
\hat{x}\hat{S}(a)-\hat{S}(a)\hat{x}=i\hbar\frac{d\hat{S}(a)}{d\hat{p}}=a\hat{S}(a).
\label{3.2}
\end{align}
Thus we obtain 
\begin{align}
\hat{x}\hat{S}(a)\vert x\rangle=(x+a)\hat{S}(a)\vert x\rangle,\quad \hat{S}(a)\vert x\rangle=\vert x+a\rangle, \qquad  \langle x_1\vert \hat{S}(a)\vert x_2\rangle=\langle x_1\vert x_2+a\rangle=\delta(x_1-x_2-a).
\label{3.3}
\end{align}
Expanding (\ref{3.3}) for small $a$ thus gives 
\begin{align}
\langle x_1\vert \left(1-\frac{ia}{\hbar}\hat{p}\right)\vert x_2\rangle=\delta(x_1-x_2)-a\frac{d}{dx_1}\delta(x_1-x_2),
\label{3.4}
\end{align}
so that
\begin{align}
\langle x_1\vert\hat{p}\vert x_2\rangle=-i\hbar\frac{d}{dx_1}\delta(x_1-x_2).
\label{3.5}
\end{align}
Thus within position operator eigenstates the momentum operator acts as a derivative even though the states are not square integrable. 

Similarly, we obtain 
\begin{align}
\langle x_1\vert\hat{p}\vert p_1\rangle=p_1\langle x_1\vert p_1\rangle=\int dx_2\langle x_1\vert\hat{p}\vert x_2\rangle\langle x_2\vert p_1\rangle=-i\hbar\frac{d}{dx_1}\langle x_1\vert p_1\rangle.
\label{3.6}
\end{align}
Thus on integrating we identify
\begin{align}
\langle x_1\vert p_1\rangle=\frac{1}{(2\pi)^{1/2}}e^{ip_1x_1/\hbar}=\psi_{1}(x_1)
\label{3.7}
\end{align}
as a plane wave,
with the normalization condition 
\begin{align}
\langle p_1\vert p_2\rangle=\int_{-\infty}^{\infty}dx \langle p_1\vert x\rangle \langle x \vert p_2\rangle=
\frac{1}{2\pi}\int_{-\infty}^{\infty}dx e^{i(p_2-p_1)x/\hbar}=\delta(p_2-p_1)
\label{3.8}
\end{align}
then following. Similarly, for position eigenstates we have  
\begin{align}
\langle x_1\vert x_2\rangle=\int_{-\infty}^{\infty}dp \langle x_1\vert p\rangle \langle p \vert x_2\rangle=\frac{1}{2\pi}\int_{-\infty}^{\infty}dp e^{i(x_1-x_2)p/\hbar}=\delta(x_1-x_2).
\label{3.9}
\end{align}
Thus even though plane waves are not square integrable,  taking matrix elements  of operators such as $\hat{p}^2/2m+V(\hat{x})$ between the bra $\langle x\vert$ and the ket $\vert \psi(x,t) \rangle$ will then lead us to the standard differential Schr\"odinger equation $i\hbar\partial_t\psi(x,t)=[-(\hbar^2/2m)\partial_x^2+V(x)]\psi(x,t)$, where the wave function $\psi(x,t)$ is given by the bra-ket overlap $\langle x\vert\psi(x,t) \rangle$, with this being the case regardless of whether or not  the $\langle \psi(x,t)\vert\psi(x,t)\rangle=\int_{-\infty}^{\infty} dx \langle \psi(x,t)\vert x\rangle\langle x\vert \psi(x,t)\rangle=\int_{-\infty}^{\infty} dx \psi^*(x,t)\psi(x,t)$ normalization  integral is finite.

Also we have
\begin{align}
&p_1\delta(p_1-p_2)=\langle p_1\vert \hat{p}\vert p_2\rangle=\int _{-\infty}^{\infty}dx_1\int _{-\infty}^{\infty}dx_2\langle p_1\vert x_1\rangle \langle x_1\vert \hat{p}\vert x_2\rangle\langle x_2\vert p_2\rangle
\nonumber\\
&=\int _{-\infty}^{\infty}dx_1\int _{-\infty}^{\infty}dx_2\psi^*_1(x_1)\left(-i\hbar\frac{d}{dx_1}\delta(x_1-x_2\right)\psi_2(x_2)
=\int _{-\infty}^{\infty}dx_1\psi^*_1(x_1)p_1\psi_2(x_1).
\label{3.10}
\end{align}
Thus  for two plane waves we  have
\begin{align}
&\int_{-\infty}^{\infty} dx \psi^*_1(x) \hat{p}\psi_2(x)- \bigg{[}\int_{-\infty}^{\infty} dx \psi^*_2(x) \hat{p}\psi_1(x)\bigg{]}^*=
\int_{-\infty}^{\infty} dx\psi^*_1(x)\psi_2(x)p_1-\int_{-\infty}^{\infty} dx\psi^*_1(x)\psi_2(x)p_2
\nonumber\\
&=
(p_1-p_2)\delta(p_2-p_1).
\label{3.11}
\end{align}
Thus formally we see that the momentum operator obeys the standard  Hermitian condition $p_{12}=p_{21}^*$ when acting on its own eigenstates.  However, this is only formal since none of the integrals in (\ref{3.11}) can be treated as ordinary integrals, and we can only set $(p_1-p_2)\delta(p_2-p_1)=0$ after integrating with a well-behaved test function, the rigged Hilbert space situation.  A similar analysis holds for the Hamiltonian. If we set $\hat{H}\vert \psi\rangle=E\vert \psi \rangle$, $\langle \psi \vert \hat{H}^{\dagger}=\langle \psi\vert E^*$, we obtain
\begin{align}
\langle \psi\vert\hat{H}\vert \psi\rangle-\langle \psi\vert\hat{H}^{\dagger}\vert \psi\rangle=(E-E^*)\langle \psi\vert\psi\rangle=(E-E^*)\int_{-\infty}^{\infty} dx \langle \psi\vert x\rangle\langle x\vert\psi\rangle=(E-E^*)\int_{-\infty}^{\infty} dx\psi^*(x)\psi(x).
\label{3.12}
\end{align}
Thus if $\psi(x)$ is not square integrable none of the matrix elements in (\ref{3.12}) exist and the connection between Hermiticity and the reality of energy eigenvalues does not hold,   this being the case for the exponentially growing radial behavior associated with the $E_0-i\Gamma$ solution to the square-well Schr\"odinger equation. (In fact one only needs $\langle \psi\vert\psi\rangle$ to be non-zero, as is the case for the radially falling $E_0+i\Gamma$ solution to the Schr\"odinger equation -- $\int_0^{\infty}e^{-2 k_2r}dr=1/2k_2$ where $k_2$ is the imaginary part of the momentum of the state.)
This is of concern for scattering by a Hamiltonian of the form  $\hat{H}=\hat{p}^2/2m+V(\hat{x})$ with a real potential, since if the potential  is short range it does not affect the large $x$ behavior of the integrals in (\ref{3.12}) as it can only  give an $x$-independent, energy-dependent $e^{i\delta}$ phase to $\psi(x)$. Consequently,  the integrals in (\ref{3.12}) do not exist as they are controlled by the asymptotic  behavior of the momentum eigenstates, non-square-integrable states that are instead formally normalized to delta functions. (For a plane wave $\langle \psi \vert \psi \rangle=\int_{-\infty}^{\infty} dx \psi^*_1(x)\psi_1(x)=\int_{-\infty}^{\infty} dx$, which is infinite.) As we now show, delta function normalization entails complex plane contours.

\section{Plane Waves and Contour Integrals}
\label{S4}

To evaluate the significance of the relation in (\ref{3.8})  we introduce the theta function and set
\begin{align}
\theta(p)=\frac{1}{2\pi i}\int _{-\infty}^{\infty}dx\frac{e^{ipx}}{x-i\epsilon}.
\label{4.1}
\end{align}
If $p$ is positive the integral is suppressed on the upper-half circle in the complex $x$ plane,  while if $p$ is negative the integral is suppressed on the lower-half circle. With a pole at $x=+i\epsilon$ being in the upper-half plane, contour integration in the complex $x$ plane then establishes that the integral in (\ref{4.1}) is indeed $\theta(p)$. (If $p=0$ the upper half plane circle at infinity contribution is not suppressed, and it serves to cancel half of the pole contribution to give $\theta(0)=1/2$.) Taking now the derivative of (\ref{4.1}) with respect to $p$ then recovers (\ref{3.8}), to thus establish that $\delta(p)$ is  given by contour integration in the complex $x$ plane, and that (\ref{3.8}) is not an ordinary integral, something it could not be since a delta function is not an ordinary function. In addition, we recall that in Sec. \ref{S1} we had discussed the role of complex Lorentz transformations in deriving the $CPT$ theorem. As we see from  (\ref{4.1}) delta functions also lead us to a role for complex coordinates. 

In analog to (\ref{4.1}), for (\ref{3.9})  we set 
\begin{align}
\theta(x)=-\frac{1}{2\pi i}\int _{-\infty}^{\infty}dp\frac{e^{-ipx}}{p+i\epsilon}.
\label{4.2}
\end{align}
If $x$ is positive the integral is suppressed on the lower-half circle in the complex $p$ plane,  while if $x$ is negative the integral is suppressed on the upper-half circle. With a pole at $p=-i\epsilon$ being in the lower-half plane, contour integration in the complex $p$ plane then establishes that  the integral in (\ref{4.2}) is indeed $\theta(x)$. Taking now the derivative of (\ref{4.2}) with respect to $x$ then recovers (\ref{3.9}), to thus establish that $\delta(x)$ is given by contour integration in the complex $p$ plane, and that (\ref{3.9}) is not an ordinary integral. The $1/(p+i\epsilon)$ factor in (\ref{4.2}) translates into the existence of a branch cut in the scattering amplitude. In addition, going from  above threshold continuum scattering states with $\psi_k(x)=e^{ikx}$ to  below threshold bound states with $\psi_{\sigma}(x)=e^{-\sigma x}$ also involves a complex plane continuation, viz. $k\rightarrow i\sigma$. Moreover, complexifying the momenta to $k=k_1+ik_2$ is a continuation that will lead us to the spatial wave functions obtained in Sec. \ref{S5}, wave functions  that are associated with the $E_0\pm i \Gamma$ energy eigenvalues that are found there. Thus all of this complex momentum plane continuation is implicit in setting $\langle x_1\vert x_2\rangle=\delta(x_1-x_2)$ and using (\ref{4.2}).

The probability for  evolution  from position $x_1$ at time $t=0$ to position $x_2$ at  positive or negative time $t$ is given by $\langle x_1\vert e^{-i\hat{H}t}\vert x_2\rangle$, to thus involve $\langle x_1\vert x_2\rangle$ since 
\begin{align}
\langle x_1\vert e^{-i\hat{H}t}\vert x_2\rangle=\langle x_1\vert x_2\rangle-it\langle x_1\vert \hat{H}\vert x_2\rangle +\ldots 
\label{4.3}
\end{align}
Thus according to (\ref{3.9})  this process also necessarily involves complex momenta simply because of the delta function normalization of $\langle x_1\vert x_2\rangle$. Similarly, if we restrict the propagation to be  forward in time  we replace $\langle x_1\vert e^{-i\hat{H}t}\vert x_2\rangle$ by $\langle x_1\vert \theta(t)e^{-i\hat{H}t}\vert x_2\rangle$, and since $\theta(t)$ can be represented by the contour integral
\begin{align}
\theta(t)=-\frac{1}{2\pi i}\int _{-\infty}^{\infty}d\omega\frac{e^{-i\omega t}}{\omega+i\epsilon},
\label{4.4}
\end{align}
time ordering causes processes to  involve complex energy. In (\ref{6.6}) and \ref{6.9}) below we  will\ show how to use (\ref{4.4}) in order to implement causality when we have both $e^{-i(E_0-i\Gamma)t}$ and $e^{-i(E_0+i\Gamma)t}$ wave functions. Also, we note that in the scattering amplitude we can continue  the energy from above to below threshold. Thus  representing theta and delta functions as contour integrals rather than as distributions enables us to readily identify the complex plane structure in both momentum and energy.

\section{The Square Well as a Worked example}
\label{S5}
\subsection{Complex conjugate pairs of scattering state solutions to the Schr\"odinger equation}
\label{S5a}
To see the role of complex momenta and energies we recall the treatment of scattering of a particle of mass $m$ by a three-dimensional, spherically-symmetric, finite-depth square well potential $V(r<a)=0$, $V(r>a)=V_0$, where $V_0$ is real, positive and  finite and $a$ is the radius of the potential. An incoming plane wave in the $z$ direction   can be expanded as a sum over angular momentum modes  of the form $e^{ikz}=\sum_{\ell=0}^{\infty}(2\ell+1)i^{\ell}j_{\ell}(kr)P_{\ell}(\cos\theta)$. The $\ell=0$ mode is spherically symmetric, with it being its interference with the other angular momentum components that 
causes the sum on $\ell$ to be in the $z$ direction. With the potential being spherically symmetric each angular momentum mode scatters separately. With $E=\hbar^2 K^2/2m$, $E-V_0=\hbar^2k^2/2m$, the radial $\ell=0$ wave functions are given by $R_0(r<a)=A \sin(Kr)/r$, $R_0(r>a)=B\sin(kr+\delta)/r$ with an energy-dependent phase shift $\delta$. Then by continuity of both the wave function and its derivative at $r=a$,  the s-wave phase shift $\delta$ acquired by the $\ell=0$ component of the outgoing wave is found to obey
\begin{align}
A \sin(Ka)=B\sin(ka+\delta), \qquad A K\cos(Ka)=Bk\cos(ka+\delta),\qquad K\tan(ka+\delta)=k\tan(Ka),
\label{5.1}
\end{align}
with the third relation holding provided $\cos(Ka)$ and $k$ are not equal to zero (which would require $A=B\sin\delta$). 
With $\beta=k\tan(Ka)/K$ and $ka=\alpha$, from the third relation in (\ref{5.1}) we obtain
\begin{align}
\sin\delta =\frac{\beta\cos\alpha-\sin\alpha}{(1+\beta^2)^{1/2}},\qquad \cos\delta =\frac{\beta\sin\alpha+\cos\alpha}{(1+\beta^2)^{1/2}}, \qquad \tan\delta=\frac{\beta-\tan\alpha}{\beta\tan\alpha+1}.
\label{5.2}
\end{align}

Bound states that  lie in the well have real $0<E<V_0$, with  incoming and outgoing scattering waves having positive, real $E>V_0$. To study the structure of solutions to (\ref{5.1}) we note that
\begin{align}
\sin(kr+\delta)=\frac{\cos\delta}{2}[e^{ikr}(\tan\delta -i)+e^{-ikr}(\tan\delta+i)].
\label{5.3a}
\end{align}
Bound states with $E<V_0$ have pure imaginary $k=i\sigma$, so that 
\begin{align}
\sin(kr+\delta)=\frac{\cos\delta}{2}[e^{-\sigma r}(\tan\delta -i)+e^{\sigma r}(\tan\delta+i)].
\label{5.4a}
\end{align}
Thus for bound states that fall off exponentially in $r$ we have $\tan\delta +i=0$. From (\ref{5.2}) it then follows that $\beta=-i$. Thus we have
\begin{align}
\beta=
\frac{ka\tan(Ka)}{Ka}=-i,\qquad \cos(Ka)=\frac{k}{(k^2-K^2)^{1/2}},\quad \sin(Ka)=-\frac{iK}{(k^2-K^2)^{1/2}},
\label{5.5a}
\end{align}
with bound state energies thus obeying 
\begin{align}
\sin(Ka)=\left(\frac{E}{V_0}\right)^{1/2}.
\label{5.6a}
\end{align}

We can immediately find bound state solutions that satisfy (\ref{5.6a})   since $\sin(Ka)$ and $E/V_0$ can both be less than one simultaneously.
However, there are other solutions to (\ref{5.6a}). We can also satisfy (\ref{5.6a}) right at the top of the well where $E=V_0$ provided $\sin(Ka)=1$, i.e., provided $V_0$ is such that $\sin[(2mV_0/\hbar^2)^{1/2}a]=1$. We discuss these particular solutions, solutions that are at exceptional points,  in Sec. \ref{S5b}. 

Above threshold there can only be solutions to (\ref{5.6a}) if $E$ and $K$ are complex. And if there are any such solutions they must appear in complex conjugate pairs since 
$\sin(Ka)=(\hbar^2/2m)^{1/2}K/V_0^{1/2}$ is a real equation (we can write it as a power series in $K$ all of whose coefficients are real). To show that there actually are such complex solutions we set $E=(\mu+i\nu)^2$,  $(2m/\hbar^2)^{1/2}a=\gamma$, $Ka=\gamma(\mu+i\nu)$, and from (\ref{5.6a})  we obtain
\begin{align}
&\sin(\gamma\mu)\cosh(\gamma\nu)=\frac{\mu}{V_0^{1/2}},\quad \cos(\gamma\mu)\sinh(\gamma\nu)=\frac{\nu}{V_0^{1/2}},
\nonumber\\
&\cosh[\gamma[\mu^2\cot^2(\gamma\mu)-V_0\cos^2(\gamma\mu)]^{1/2}]=\frac{\mu}{V_0^{1/2}\sin(\gamma\mu)},\qquad 
\cos[\gamma[\nu^2\coth^2(\gamma\nu)-V_0\cosh^2(\gamma\nu)]^{1/2}]=\frac{\nu}{V_0^{1/2}\sinh(\gamma\nu)}.
\label{5.7a}
\end{align}
From (\ref{5.7a})  it follows that
\begin{align}
\frac{\tan(\gamma \mu)}{\gamma\mu}=\frac{\tanh(\gamma \nu)}{\gamma \nu},
\label{5.8a}
\end{align}
provided that  $\nu\neq 0$.
The relations given in (\ref{5.7a}) and (\ref{5.8a}) can then be solved for the two variables $\mu$ and $\nu$. Noting that these relations  are invariant under $\nu\rightarrow -\nu$, this then gives not one but two solutions for $E$ of the form 
\begin{align}
E_0\pm i\Gamma=\mu^2-\nu^2\pm 2i\mu\nu, \qquad  \mu^2=\frac{(E_0^2+\Gamma^2)^{1/2}+E_0}{2}, \qquad \nu^2=\frac{(E_0^2+\Gamma^2)^{1/2}-E_0}{2},
\label{5.9a}
\end{align}
solutions that are not just complex but in a complex conjugate pair. We note that $V_0$ is real and we have not given the potential any imaginary part.  The existence of pairs of exact solutions to the above threshold Schr\"odinger equation does not appear to have been considered in discussions of scattering off a square well, and constitutes a main result of this paper. Also we note that with (\ref{5.6a}) we are being forced into the complex energy plane above threshold. While this is of course standard for the scattering amplitude $f(E)$ that we discuss below in Sec. \ref{S5c}, we see here that the standard analytic structure of the scattering amplitude originates in the structure of solutions to  the Schr\"odinger equation.

In regard to these solutions  we note that  for $\gamma\nu$ positive the quantity $\tanh(\gamma \nu)/\gamma \nu $ always lies between one and zero.  For $0<\gamma\mu<\pi/2$ the quantity  $\tan(\gamma \mu)/\gamma\mu $ is always  greater than one. For $\pi/2<\gamma\mu<\pi$ the quantity $\tan(\gamma \mu)/\gamma\mu $ is always negative. And since for $\pi<\gamma\mu<3\pi/2$ the quantity  $\tan(\gamma \mu)/\gamma\mu $ is not only positive but varies continuously between zero and infinity, there must be a non-trivial $\nu \neq 0$ solution to (\ref{5.8a}) (the first such one)  in the range $\pi < \gamma \mu <3\pi/2$, to thus be well above the threshold at $E=V_0$.  The next solution lies in the region $2\pi < \gamma \mu <5\pi/2$, and so on. Since $\tanh(\gamma \nu)/\gamma \nu $  always lies between one and zero for either sign of $\nu$ the solutions appear in pairs. We thus confirm that the above scattering threshold there are solutions to the square well Schr\"odinger equation that appear in complex conjugate pairs. 

Representing the complex pair of energies as $E_0\pm i\Gamma$, the wave function time dependence is of the form $\psi_{\pm}(t)=e^{-i(E_0\pm i\Gamma)t/\hbar}$, to thus be eigenstates of $i\hbar \partial_t$ even though they are not stationary. To determine the spatial behavior, we note that with $\tan\delta =-i$ the wave functions are  given by 
\begin{align}
\psi_{\pm}(r<a,t)=A\frac{\sin((K_1\pm iK_2)r)}{r}e^{-iE_0t\pm \Gamma t}, \qquad \psi_{\pm}(r>a,t)=C\frac{e^{ik_1r\mp k_2r}}{r}e^{-iE_0t\pm \Gamma t},
\label{5.10a}
\end{align}
where we have introduced $C=-iB\cos\delta$ and set $K=K_1\pm iK_2$, $k=k_1\pm ik_2$, so that 
\begin{align}
\frac{\hbar^2}{2m}(K_1^2-K_2^2)=E_0,\qquad \frac{\hbar^2K_1K_2}{m}=\Gamma,\qquad \frac{\hbar^2}{2m}(k_1^2-k_2^2)=E_0-V_0,\qquad \frac{\hbar^2k_1k_2}{m}=\Gamma.
\label{5.11a}
\end{align}
Satisfying the boundary conditions at $r=a$ leads to 
\begin{align}
\tan((K_1\pm iK_2)a)=\frac{K_1\pm iK_2}{i(k_1\pm i k_2)},\qquad \sin((K_1\pm iK_2)a)=\frac{K_1\pm iK_2}{[(K_1\pm iK_2)^2-(k_1\pm ik_2)^2]^{1/2}}=\left(\frac{E_0\pm i\Gamma}{V_0}\right)^{1/2},
\label{5.12a}
\end{align}
which we recognize as (\ref{5.6a}) as evaluated above threshold.

There is one key difference between the bound state and scattering state solutions that we have found.  For the bound states we can take the coefficients $A$ and $B$ that are given in (\ref{5.1})  to be finite. However, in the scattering sector since $\tan\delta =-i$ it follows that $\cos\delta$ is infinite. We must thus take $B$ to behave as $1/\cos\delta$ so that $C$ and then $A$ are finite. There is thus a discontinuity in the parameter $B$ as we continue from the bound sector to the scattering sector. Such a discontinuity can even be expected since while the energy eigenvalues are real below threshold there are complex conjugate pairs of energy eigenvalues above threshold. The Hamiltonian is self-adjoint when it acts in the bound state sector but is not self-adjoint when it acts on the $E_0-i\Gamma$ scattering sector energy eigenfunction because of  the exponential growth of the associated radial eigenfunction. 

If we define the  probability amplitude as the standard $\psi_{\pm}(r,t)^*\psi_{\pm}(r,t)$, the general structure of $\psi_{\pm}(r>a,t)$ would lead to a probability amplitude for $\psi_+(r,t)$ that grows exponentially in time, and a probability amplitude for $\psi_-(r>a,t)$ that grows exponentially in space. However, above threshold we cannot use the standard Dirac norm as the eigenfunctions are not square integrable, but as we describe below in Sec. \ref{S5e},  we  are  in a $PT$ theory. And a $PT$ theory has a different probability amplitude, the one associated with the $\hat{V}$ operator that we discuss in Sec. \ref{S6}. Its structure  is such that the only non-zero probability amplitudes are of a time-independent form 
\begin{align}
&\psi_{\pm}(r<a,t)^*\psi_{\mp}(r<a,t)=\mp\frac{A^*\sin((K_1\mp iK_2)r)A\sin((K_1\mp iK_2)r)e^{iE_0t\pm\Gamma t-iE_0t\mp\Gamma t}}{r^2}=\mp\frac{A^*A\sin^2((K_1\mp iK_2)r)}{r^2},
\nonumber\\
&\psi_{\pm}(r>a,t)^*\psi_{\mp}(r>a,t)=\mp C^*C\frac{e^{-ik_1r\mp k_2r+iE_0t\pm\Gamma t}}{r}\frac{e^{ik_1r\pm k_2r-iE_0t\mp\Gamma t}}{r}=\mp\frac{C^*C}{r^2}
\label{5.13a}
\end{align}
that is well behaved at both large time and large space, with all exponentially growing terms being  cancelled by compensating exponentially falling ones. Since $(e^{-ik_1r}/r)(e^{ik_1r}/r)$ vis also equal to $1/r^2$, in $r>a$ the probability amplitude given in (\ref{5.13a}) is the same as the standard one associated with free delta-function normalized spherical  waves.

\subsection{Exceptional points}
\label{S5b}

Moreover, it is also of interest to note that not only do we get complex conjugate pairs of eigenvalues when $\nu\neq 0$, in the  $\nu=0$ limit the two energy eigenvalues in (\ref{5.9a}) collapse into just one common eigenvalue, viz. that with $E=\mu^2$. To see what happened to  the second solution, we first discuss the common eigenvalue case itself, and note that when $E=\mu^2$ and $\nu=0$, viz. $\mu^2=K^2a^2/\gamma^2=\hbar^2K^2/2m=E$, there will be a solution to (\ref{5.6a}) if $\sin(\gamma E^{1/2})=(E/V_0)^{1/2} $. And it is possible to have a solution when $E$ is at the scattering threshold, i.e.,  $E=V_0$, though for this to happen  $V_0$ has to obey $\sin(\gamma V_0^{1/2})=1$, i.e., obey any of the conditions  
\begin{align}
&\gamma V_0^{1/2}=\frac{\pi}{2},~~\frac{3\pi}{2},~~\frac{5\pi}{2},\ldots \qquad V_0=\frac{\pi^2\hbar^2}{8ma^2}, ~~ \frac{9\pi^2\hbar^2}{8ma^2},~~~\frac{25\pi^2\hbar^2}{8ma^2},\ldots
\label{5.14a}
\end{align}
These conditions require very specific values for $V_0$. We recall that for the square well we need $V_0>\pi^2\hbar^2/8m a^2$ in order to form the first square-well bound state, and then $V_0>9\pi^2\hbar^2/8m a^2$ in order to form a second bound state, and so on. Thus when $V_0=\pi^2\hbar^2/8m a^2$,  $V_0=9\pi^2\hbar^2/8m a^2$ and so on, bound states can just be formed right at the top of the well. Since these modes lie right at the scattering threshold they satisfy the second equation in (\ref{5.1}) by having both $\cos(Ka)$ and $k$ vanish, i.e., by having $\sin{\gamma \mu}$ and $E-V_0$ vanish. 

For states right at the top of the well with $E=V_0$, $k=0$, $\sigma=0$, 
the relevant wave function is based on $1/r$ (the $k=0$ limit of $e^{ikr}/r$), and is found to be of the form 
\begin{align}
&\psi_1(r<a,t)=\frac{A_1e^{-iV_0t/\hbar}\sin((2mV_0/\hbar^2)^{1/2}r)}{r},\qquad \psi_1(r>a,t)=\frac{B_1e^{-iV_0t/\hbar}}{r},\qquad A_1\sin((2mV_0/\hbar^2)^{1/2}a)=B_1,
\nonumber\\
&\cos((2mV_0/\hbar^2)^{1/2}a)=0,\qquad V_0=\frac{\pi^2\hbar^2}{8ma^2}, ~~ \frac{9\pi^2\hbar^2}{8ma^2},~~~\frac{25\pi^2\hbar^2}{8ma^2},\ldots,
\label{5.15a}
\end{align}
On  noting that for $r>a$ $\nabla^2(1/r)=(\partial_r^2+(2/r)\partial_r)(1/r)=0$,  we can show directly that $\psi_1(r,t)$ obeys the Schr\"odinger equation both inside and outside the well, obeys the boundary conditions on $\psi_1(r,t)$ and $\partial_r\psi_1(r,t)$ at $r=a$, is an eigenstate of $i\hbar\partial_t$ with eigenvalue $V_0$, and obeys $\tan\delta+i=0$. The state is not bound and lies right at the  beginning of the continuum, with the $r>a$ contribution to the normalization integral $\int_a^{\infty}4\pi r^2 dr \psi_1^*\psi_1=(\int_0^{\infty}-\int_0^{a})4\pi r^2 dr \psi_1^*\psi_1$ being dominated by $4\pi\int_0^{\infty}dr$, to thus have a leading term $4\pi^2 \delta(0)$ ($\pi \delta(0)$ is the $k \rightarrow 0$ limit of $\int_0^{\infty} dr e^{ikr}$), the standard plane wave normalization.

In addition, there is a second solution, one with a radically different structure. We recall that in Sec. \ref{S1} we had noted that in the limit $\Gamma \rightarrow 0$ we have $[e^{-i(E_0-i\Gamma) t}+e^{-i(E_0+i\Gamma) t}]/2\rightarrow e^{-iE_{0} t}$, $[e^{-i(E_0-i\Gamma) t}-e^{-i(E_0+i\Gamma) t}]/(-2\Gamma) \rightarrow te^{-iE_{0} t}$, to give one solution, $\psi_1(r,t)$,  that is stationary, and a second one, $\psi_2(r,t)$,  that grows linearly in time. We thus seek such a linearly growing  in time solution for the square well as it has such a complex conjugate pair solution. And on trying as a generic form  $\psi_2(x,t)=e^{-iV_0 t/\hbar}[c(r)t+d(r)]$ with different $c(r)$ and $d(r)$ inside and outside the well, we find the solution to be of the form:
\begin{align}
&\psi_2(r<a,t)=A_2e^{-iV_0t/\hbar}\left(\frac{\sin((2mV_0/\hbar^2)^{1/2}r)t}{r}+\frac{im\cos((2mV_0/\hbar^2)^{1/2}r)}{(2mV_0)^{1/2}}\right),
\nonumber\\
&\psi_2(r>a,t)=B_2e^{-iV_0t/\hbar}\left(\frac{t}{r}-\frac{im(r-a)}{\hbar}\right),\qquad A_2\sin((2mV_0/\hbar^2)^{1/2}a)=B_2,\qquad \cos((2mV_0/\hbar^2)^{1/2}a)=0,
\nonumber\\
& V_0=\frac{\pi^2\hbar^2}{8ma^2}, ~~ \frac{9\pi^2\hbar^2}{8ma^2},~~~\frac{25\pi^2\hbar^2}{8ma^2},\ldots,
\label{5.16a}
\end{align}
where again $V_0$ has to take the specific values given in (\ref{5.14a}), even though this time the Schr\"odinger equation solutions given in (\ref{5.16a}) are not energy eigenstates.
The linear in $t$ dependence of $\psi_2(r,t)$  is given by $\psi_1(r,t)$ as multiplied by $t$, and the additional terms are the ones needed to balance the effect of $i\hbar \partial_t$ on the linear term in $t$ and meet the boundary conditions at $r=a$.  As such, $\psi_2(r,t)$ obeys the Schr\"odinger equation both inside and outside the well, obeys the boundary conditions on $\psi_2(r,t)$ and $\partial_r\psi_2(r,t)$ at $r=a$, but is not an eigenstate of $i\hbar\partial_t$, and does not obey $\tan\delta+i=0$. Apart from already being of interest in as much as a solution of this form even exists, this linear in time growth is typical of exceptional points. Exceptional points are characteristic of systems at which a Hamiltonian loses an eigenstate and  no longer has a complete set of eigenstates, to thereby  become of non-diagonalizable, Jordan-block form, with the eigenstate that is lost becoming a non-stationary solution to the Schr\"odinger equation  that grows linearly (or more generally as a power) in time (see e.g. \cite{Mannheim2018a} and references therein). 

Like $\psi_1(r,t)$ we find that $\psi_2(r,t)$ is also continuum normalized, viz.
\begin{align}
\int_a^{\infty}4\pi r^2 dr \psi_2^*(r>a,t)\psi_2(r>a,t)&=4\pi \int_a^{\infty}r^2dr\left(\frac{t^2}{r^2}+\frac{m^2(r-a)^2}{\hbar^2}\right)
\nonumber\\
&\sim 4\pi^2\left[t^2\delta(0)+\frac{m^2}{\hbar^2}\left(\delta^{\prime\prime\prime\prime}(0)+2a^2\delta^{\prime\prime}(0)+a^2\delta(0)\right)\right],
\label{5.17a}
\end{align}
though more singularly so than $\psi_1(r,t)$.  Despite this singular structure, not only are $\psi_1(r,t)$ and $\psi_2(r,t)$ solutions to the theory when $E=V_0$ and $V_0=\pi^2\hbar^2/8ma^2,~ 9\pi^2\hbar^2/8ma^2,~25\pi^2\hbar^2/8ma^2,\ldots$, they are the only such solutions. They are quite different from the $E<V_0$ bound state solutions (ones that behave as $e^{-\sigma r}/r$), solutions for which there is no linear in $t$ counterpart, with only the threshold branch point being able to be an exceptional point, and even then  only when $V_0=\pi^2\hbar^2/8ma^2,~ 9\pi^2\hbar^2/8ma^2,~25\pi^2\hbar^2/8ma^2,\ldots$ Thus both $\psi_1(r,t)$ and $\psi_2(r,t)$ belong to the scattering sector, sitting right where it begins at threshold, with the threshold branch point thus being an exceptional point.

This linear in time behavior is quite generic. In \cite{Mannheim2026} we have found an analogous behavior for both the simple harmonic oscillator and for the free theory. For the free theory with $i\hbar\partial_t\psi(x,t)=-(\hbar^2/2m)\partial_x^2\psi(x,t)$ we find that $\psi(x,t)=x$ and $\psi(x,t)=tx-imx^3/3\hbar$ are both exact solutions, with the first solution being an energy eigenstate with $E=0$ (viz. right at the scattering threshold, just as in the square well case), and with  the second one not being an energy  eigenstate at all because of its linear in time growth.  In this sense the square well solutions given in (\ref{5.15a}) and (\ref{5.16a}) actually inherit that structure from the same structure in the free theory, its plane wave eigenfunctions not being square integrable. Finally, in regard to probability conservation, using the $\hat{V}$ operator procedure described in Sec. \ref{S6} we can construct a time-independent probability amplitude for  the simple harmonic oscillator, the square well,  and the free theory even though there is linear in time growth in all three cases. We discuss  these issues in detail in  \cite{Mannheim2026}.

\subsection{Scattering amplitude}
\label{S5c}

In discussing the scattering amplitude there are two approaches, the Breit-Wigner one and the $PT$ one. For the standard Breit-Wigner approach one uses the square well Schr\"odinger equation given in (\ref{5.1})  as evaluated at real $E$, and in terms of the phase shift $\delta$ constructs a real $E$ scattering amplitude of the form 
\begin{align}
f_{\rm BW}(E)=e^{i\delta}\sin\delta= \frac{\tan\delta}{1-i\tan\delta}=\frac{(\beta-\tan\alpha)}{(\beta+i)(\tan\alpha-i)}=\frac{ka\tan(Ka)-Ka\tan\alpha}{(ka\tan(Ka)+iKa)(\tan\alpha-i)}.
\label{5.18a}
\end{align}
According to (\ref{5.2}) the phase shift goes through $\pi/2$ (or an odd multiple thereof) when
\begin{align}
\beta=
\frac{ka\tan(Ka)}{Ka}=-\frac{1}{\tan(ka)}.
\label{5.19a}
\end{align}
Denoting the resonance value of $E$ as  $E_0$, then near resonance  we can set
\begin{align}
\beta \tan(ka)+1=E_0-E, \qquad \tan(ka)=\frac{(E_0-E-1)}{\beta},
\label{5.20a}
\end{align}
as momentarily written in dimensionless units.
Denoting by  $\beta_0$  the value of $\beta$ at $E=E_0$, using (\ref{5.18a}) near resonance we can set
\begin{align}
\tan\delta&=\frac{\Gamma}{E_0-E},\qquad \Gamma=\beta_0 +\frac{1}{\beta_0},\qquad f_{\rm BW}(E)=\frac{\Gamma}{E_0-i\Gamma-E},\nonumber\\
 \vert f_{\rm BW}(E)\vert^2&=\frac{\Gamma^2}{(E_0-E)^2+\Gamma^2},\qquad \vert f_{\rm BW}(E_0)\vert^2=1,
\label{5.21a}
\end{align}
just as given in (\ref{2.1}). The sign of $\Gamma$ would then be determined by the sign of $\beta_0$, and there could be $\tan(\delta)=\infty$ cases in which $\beta_0$ and thus 
$\Gamma$ would be negative rather than positive.

While a subsequent continuation of the function $f_{\rm BW}(E)$ into the complex energy plane can be made, the pole that would be obtained at $E=E_0-i\Gamma$ would not be physical unless it just happened to be associated with an eigenstate of the Hamiltonian and obey (\ref{5.7a}) and $\tan\delta(E_0-i\Gamma)=-i$. (We have checked numerically in a typical case that this is not so, even finding a negative value for $\beta_0$ and thus for $\Gamma$ in it \cite{footnote00}). However, resonances in a scattering cross-section  $\vert f_{\rm BW}(E)\vert^2$ are identified with physical particles, especially in high energy scattering processes. For such particles to be physical we must be able to identify them with solutions to the Schr\"odinger equation. To this end we thus turn to the $PT$ symmetry approach.

In Sec. \ref{S5a} we found a complex conjugate pair of energy eigenvalue solutions to the Schr\"odinger equation  at $E_0\pm i\Gamma$. Both solutions obeyed of $\tan \delta_{\mp}=-i$, to thus be poles in the scattering amplitude. To make contact with resonances in the real $E$ scattering amplitude we introduce phase shifts $\delta_-(E)$ and $\delta_+(E)$ of the form
\begin{align}
\tan\delta_-(E)=\frac{\Gamma}{E_0-E},\qquad \tan\delta_+(E)=-\frac{\Gamma}{E_0-E}.
\label{5.22}
\end{align}
At the poles they obey $\tan\delta_-(E_0-i\Gamma)=-i$, $\tan\delta_+(E_0+i\Gamma)=-i$, just as poles should. At real $E=E_0$ they obey $\delta_-(E_0)=\pi/2$, $\delta_+(E_0)=-\pi/2$, just as resonances should. Together, they give a scattering amplitude of the form 
\begin{align}
f_{\rm PT}(E)&=\frac{\tan\delta_-(E)}{1-i\tan\delta_-(E)}+\frac{\tan\delta_+(E)}{1-i\tan\delta_+(E)}=\frac{\Gamma}{E_0-i\Gamma-E}-\frac{\Gamma}{E_0+i\Gamma-E}=\frac{2i\Gamma^2}{(E-E_0)^2+\Gamma^2}, 
\nonumber\\
\vert f_{\rm PT}(E) \vert^2&=\frac{4\Gamma^4}{[(E-E_0)^2+\Gamma^2]^2},\qquad \vert f_{\rm PT}(E_0) \vert^2=4.
\label{5.23}
\end{align}

Comparing with  $f_{\rm BW}(E)$, we see that $f_{\rm BW}(E)$ is derived at  real $E$ and continued to complex $E$, while in contrast $f_{PT}(E)$ is derived at the complex $E$ poles and continued to real $E$. The phase shift in $f_{\rm BW}(E)$ obeys the real $E$ (\ref{5.1}), while the phase shifts in $f_{PT}(E)$ are built from solutions that obey (\ref{5.7a}) at the complex poles. The real $E$ solution to (\ref{5.1}) and the complex $E$ solution to (\ref{5.7a}) do not continue into each other as $E$ is continued, since for that to happen both would have to remain solutions to the Schr\"odinger equation at every intermediate $E$, and as we had noted above, we had checked numerically in a specific case that this is not the case. Thus despite the notation the $E_0$ and $\Gamma$ determined from (\ref{5.1}) and (\ref{5.7a}) are different, with it being the ones determined from (\ref{5.7a}) that are the relevant ones for particles that are found in resonant scattering as they have to be dynamical eigenstates of the scattering Hamiltonian \cite{footnote00}. Since scattering cross sections are determined experimentally, identifying the associated  $E_0$ and $\Gamma$ is actually an identification  using $f_{PT}(E)$ even though it was thought to be one using $f_{\rm BW}(E)$. Fitting using $f_{PT}(E)$ or $f_{\rm BW}(E)$ would lead to the same extracted $E_0$  parameter as the two amplitudes behave the same way on resonance at $E=E_0$, though they could in principle be distinguished off resonance and give different $\Gamma$ determinations. The presence of resonances in the real energy scattering cross-section thus indicates that the Hamiltonian is not Hermitian (the complex poles are eigenstates), and that the continuation not of $D_{\rm BW}(E)$ but of $D_{ PT}(E)$ between real $E$ and the complex poles is both needed and valid. 

The contrast here with the standard Breit-Wigner discussion can be stated as follows. For the Breit-Wigner case one starts with a real energy resonance and continues $f_{\rm BW}(E)$ into the complex energy plane, with there being no complex energy eigenstate. In the $PT$ case described here one instead starts with the Schr\"odinger equation, obtains bona fide complex conjugate pairs of energy eigenstates, and then continues the $f_{PT}(E)$ scattering amplitude from the associated complex $E$ plane poles to real $E$ and obtains bona fide resonances. This is of significance since high energy scattering cross section resonances are identified with actual physical particles, and thus cannot be associated with a purely mathematical continuation of the scattering amplitude. Rather, a posteriori, we see that they should be associated with not one complex pole but with a complex conjugate pair of poles just as antilinear symmetry requires.

In addition we note that the modes that grow linearly  in time  that we have found have no scattering amplitude counterpart, to thus show that while the scattering amplitude captures the structure of the energy eigenstates of the Schr\"odinger equation, it is unable to capture solutions that are not energy eigenstates. Since we have found exceptional point exact solutions to the Schr\"odinger equation that are not energy eigenstates at all, the Hamiltonian is not diagonalizable and thus could not be Hermitian. The complex conjugate pair solutions to the square-well Hamiltonian involve solutions some of which are  not square integrable. In Sec. \ref{S6} we show that in the complex conjugate pair solution the Hamiltonian is pseudo-Hermitian, and in \cite{Mannheim2026} we show that this is also the case in the exceptional point situation. Pseudo-Hermiticity or analogously $PT$ symmetry is thus the general description of the square well system.

\subsection{One-dimensional square well}
\label{S5d}

A situation similar to that of the three-dimensional square well is met in the one-dimensional case, where $V(x<-a/2)=V_0$, $V(-a/2<x<a/2)=0$, $V(x>a/2)=V_0$, with $V_0$ being real and positive. With incoming wave $Ae^{ikx}$, reflected wave $Be^{-ikx}$, a wave in the well region of the form $Fe^{iKx}+Ge^{-iKx}$,  and transmitted  wave $Ce^{ikx}$, by continuity of  the wave function and its derivative at both $x=-a/2$ and $x=a/2$ we have
\begin{align}
\frac{C}{A}=\frac{e^{-ika}}{\cos(Ka)-i\sin(Ka)(k^2+K^2)/2kK}, ~~ \frac{B}{C}=\frac{i\sin(Ka)(K^2-k^2)}{2kK},~~ K=\left(\frac{2mE}{\hbar^2}\right)^{1/2},~~ k=\left(\frac{2m(E-V_0)}{\hbar^2}\right)^{1/2},
\label{5.24}
\end{align}
with the transmission $T$ and reflection $R$ coefficients being given by 
\begin{align}
T=\frac{CC^*}{AA^*}=\frac{1}{1+X^2}, \quad R=\frac{BB^*}{AA^*}=1-T=\frac{X^2}{1+X^2}, \quad X^2=\frac{(k^2-K^2)^2}{4k^2K^2}\sin^2(Ka)=\frac{V_0^2}{4E(E-V_0)}\sin^2(Ka).
\label{5.25}
\end{align}
Resonances occur in $T$ when $X=0$, viz. $\sin(Ka)=0$, at which such points $R=0$. In the complex energy plane we have poles in $C/A$ when  
\begin{align}
\tan(Ka)=-\frac{2ikK}{k^2+K^2}=\frac{2i[E(E-V_0)]^{1/2}}{2E+V_0}=\tan\left(\frac{(2mE)^{1/2}a}{\hbar}\right).
\label{5.26}
\end{align}
Just as in  the three-dimensional case, (\ref{5.26}) has complex conjugate pair solutions in the scattering sector.

\subsection{General features of the solutions}
\label{S5e}
As we see, despite that fact that $V_0$ is real, and despite the fact that the set of all plane waves is complete, in the scattering sector the Hamiltonian is acting on states that are not square integrable, so that the Hamiltonian is not self-adjoint when it acts on them.  In the scattering sector the set of all plane wave states does form a vector space, but one that does not have a square-integrable inner product. While there is a breakdown of the connection between Hermiticity and real eigenvalues in the scattering sector, and while that then does permit complex eigenvalues, that does not  in and of itself exclude the possibility that the eigenvalues might still be real. (And indeed even though its eigenfunctions are delta function normalized the position operator does have real eigenvalues.) That there actually are complex scattering sector eigenvalues is because above threshold solutions with  $\tan\delta=-i$ require complex $\delta$. 

Even if  $\tan\delta=-i$, that  does not in and of itself oblige there to be sets of pairs of complex conjugate solutions, as there could in principle be isolated complex solutions. However, on general grounds (probability conservation and  invariance under the complex  Lorentz group) it has been shown \cite{Mannheim2018a} that all processes that can be realized in nature, including the non-relativistic limit of them,  must  admit of a particular antilinear symmetry, namely $CPT$ symmetry. One can of course discuss systems  without such a symmetry (and it can be instructive to do so), but they cannot be realized in nature and will not be considered  here.

Moreover, for 
charge-conjugation-conserving processes $CPT$ defaults to the much-studied $PT$ symmetry (see e.g. \cite{Bender1998,Bender1999,Mostafazadeh2002,Bender2002,Bender2007,Makris2008,Bender2008a,Bender2008b,Guo2009,Bender2010,Special2012,Theme2013,ElGanainy2018,Bender2018,Fring2021}, \cite{Mannheim2018a}, together with  more that 13,000 other peer-reviewed published papers to date). We can always ascribe a notion of a symmetry to an operator because a symmetry is a property of the operators themselves regardless of which states they might operate on.  The significance of antilinear symmetry for our purposes here is that if  we introduce some general antilinear operator $\hat{A}$ that commutes with a Hamiltonian $\hat{H}$, and consider eigenvalues $E$ and eigenfunctions $e^{-iEt}\vert \phi\rangle$ of $\hat{H}$ that obey $\hat{H}\vert \phi \rangle=E\vert \phi\rangle$ we obtain
\begin{align}
\hat{H}\hat{A}\vert \phi\rangle=\hat{A}\hat{H}\vert \phi\rangle=\hat{A}E\vert \phi\rangle=E^*\hat{A}\vert \phi\rangle.
\label{5.27}
\end{align}
Thus for every eigenvalue $E$ with eigenvector $\vert \phi\rangle$ there is an eigenvalue $E^*$ with eigenvector $\hat{A}\vert \phi\rangle$. Thus as first noted by Wigner in his study of time reversal invariance \cite{Wigner1960}, antilinear symmetry implies that energy eigenvalues are either real or in complex conjugate pairs. With the square well Hamiltonian being $PT$ symmetric (it is actually both $P$ and $T$ symmetric),  for bound states the energies are real while for scattering there are complex conjugate pairs of energies, with our example of the charge-conjugation-conserving  square well thus exhibiting these aspects of $PT$ symmetry. 

If we do not use a rigged Hilbert space the plane wave continuum wave functions would not be square integrable. Thus we cannot use Hermiticity to show that their energy eigenvalues are real. However, we can use $PT$ symmetry, And thus we should regard $PT$ symmetry as the reason continuum modes have real energy.

As we see,  with antilinear symmetry, if there is one complex solution at all there there must be a second solution, the complex conjugate one, just as we found in (\ref{5.9a}). We now show that nonetheless, even with a complex conjugate pair of complex poles in the scattering amplitude there is only one observable resonance, with it being probability conservation that requires the presence of the complex conjugate pair.

\section{Two Poles in the Scattering Amplitude but only one Resonance}
\label{S6}
While we cannot ascribe a notion of Hermiticity in the scattering sector, we can nonetheless still define a notion of Hermitian adjoint, since for every ket vector $\vert \psi\rangle$, even delta function normalized ones, we can define a dual space bra vector as $\langle \psi\vert$, just as we did in (\ref{3.1}). And thus for the ket vector $\hat{H}\vert \psi\rangle$,  we can define a dual space bra vector as $\langle \psi\vert \hat{H}^{\dagger}$, with this serving as the definition of $\hat{H}^{\dagger}$. We now introduce a time-independent operator $\hat{V}$, and with $i\partial_t\vert \psi\rangle=\hat{H}\vert \psi\rangle$, $-i\partial_t\langle \psi \vert=\langle \psi \vert \hat{H}^{\dagger}$, we evaluate
\begin{align}
i\frac{\partial}{\partial t}\langle \psi(t)\vert \hat{V}\vert \psi(t)\rangle=\langle \psi(t)\vert (\hat{V}\hat{H}-\hat{H}^{\dagger}\hat{V})\vert \psi(t)\rangle.
\label{6.1}
\end{align}
Thus $\langle \psi(t)\vert \hat{V}\vert \psi(t)\rangle$ will be time independent and probability will be conserved 
 if there exists a $\hat{V}$ that obeys 
\begin{align}
\hat{V}\hat{H}=\hat{H}^{\dagger}\hat{V},\quad \hat{V}\hat{H }\hat{V}^{-1}=\hat{H}^{\dagger},
\label{6.2}
\end{align}
with the second condition requiring that $\hat{V}$ be invertible, something we take to be the case here. The $\hat{V}\hat{H}\hat{V}^{-1}=\hat{H}^{\dagger}$ condition is  known as pseudo-Hermiticity and implements probability conservation. Pseudo-Hermiticity was introduced by Dirac \cite{Dirac1942} and Pauli \cite{Pauli1943} in their study of indefinite metric quantum field theories (a recent discussion of which may be found in \cite{Mannheim2024}), and was connected to $PT$ symmetry in \cite{Mostafazadeh2002}.
With $\hat{V}^{-1}e^{i\hat{H}^{\dagger}t}\hat{V}=e^{i\hat{H}t}$, the $\hat{V}^{-1}\hat{U}^{\dagger}(t)\hat{V}\hat{U}(t)=I$ condition  with $\hat{U}(t)= e^{-i\hat{H}t} $ is known as pseudounitarity. Pseudounitarity thus generalizes ordinary $\hat{U}^{\dagger}(t)\hat{U}(t)=I$ time evolution unitarity. While pseudo-Hermiticity generalizes standard Hermiticity (which is recovered if $\hat{V}=I$), Hermitian theories do have antilinear symmetry since historically the $CPT$ theorem was  expressly proven for Hermitian theories. In \cite{Mannheim2018a} the $CPT$ theorem  was generalized to the non-Hermitian case. Thus antilinearity does not replace Hermiticity. Rather it generalizes it so as to include cases such as complex conjugate energy eigenvalue pairs that cannot be achieved in a Schr\"odinger equation with a Hermitian Hamiltonian.  

The $\hat{V}\hat{H}\hat{V}^{-1}=\hat{H}^{\dagger}$ condition entails that the relation between $\hat{H}$ and $\hat{H}^{\dagger}$  is isospectral. Thus every eigenvalue of $\hat{H}$ is an eigenvalue of $\hat{H}^{\dagger}$. Consequently the eigenvalues of $\hat{H}$ are  either real or in complex conjugate pairs. But this is the antilinear symmetry condition. Thus for any $\hat{H}$ whose eigenspectrum is complete there always will be a $\hat{V}$ operator if $\hat{H}$ has an antilinear symmetry \cite{Mannheim2018a}, with $\langle \psi(t)\vert \hat{V}\vert \psi(t)\rangle$ being the most general inner product that one could introduce that is probability conserving. If we introduce right-eigenvectors $\vert R\rangle$ of $\hat{H}$ according to  $\hat{H}\vert R\rangle=E\vert R \rangle$, we have
\begin{align}
\langle R\vert \hat{H}^{\dagger}=E^*\langle R\vert=\langle R\vert \hat{V}\hat{H }\hat{V}^{-1},\qquad \langle R\vert \hat{V}\hat{H }=E^*\langle R\vert \hat{V}.
\label{6.3}
\end{align}
Consequently, we  can identify a left eigenvector $\langle L\vert=\langle R\vert \hat{V}$, and can write the inner product as $\langle R\vert \hat{V}\vert R\rangle=\langle L\vert R\rangle$. Thus in general we can  identify the left-right inner product as the most general probability-conserving inner product in the antilinear case, a form that could perhaps be  anticipated since a Hamiltonian cannot have any more eigenvectors than its left and right ones. Thus the way to generalize standard Hermitian quantum mechanics is to replace the dual space that is built out of the Hermitian conjugate bras $\langle R\vert$ of a given set of $\vert R\rangle$ kets by a dual space built on  $\langle R\vert \hat{V}=\langle L\vert$ bras instead. That we are able to do this is because the Schr\"odinger equation $\hat{H}\vert R\rangle=E\vert R \rangle$ only specifies the ket, leaving the bra undetermined. In general  the bra $\langle L\vert$ is not the Hermitian conjugate of $\vert R \rangle$, but is instead $\langle R\vert \hat{V}$, or equivalently the $CPT$ conjugate of $\vert R\rangle$. (Like $CPT$ conjugation Hermitian conjugation also involves complex conjugation, with $CPT$ conjugation thus being its natural generalization.)

\subsection{Probability conservation}
\label{S6a}
To illustrate the role of $\hat{V}$ we combine the two complex conjugate eigenvalues $E_0\pm i\Gamma$ of each pair into a diagonal  two-dimensional matrix of the form 
\begin{align}
M=\begin{pmatrix}E_0+i\Gamma&0\\ 0&E_0-i\Gamma \end{pmatrix}.
\label{6.4}
\end{align}
As constructed, the matrix $M$  is $PT$ symmetric under $P=\sigma_1$, $T=K$, where $K$ denotes complex conjugation, to thus naturally have complex conjugate eigenvalues. The operator  $V=-i\sigma_2$ effects $VMV^{-1}=M^{\dagger}$. With this $V$  the eigenkets, eigenbras, and the  orthogonality and closure relations associated with  $M$ are given by \cite{Mannheim2018a}
\begin{align}
&u_+=e^{-iE_0t+\Gamma t}\begin{pmatrix}1\\ 0\end{pmatrix}, \qquad u_-=e^{-iE_0t-\Gamma t}\begin{pmatrix}0\\ 1\end{pmatrix}, \qquad u_+^{\dagger}V=e^{iE_0t+\Gamma t}\begin{pmatrix}0&-1 \end{pmatrix},\qquad u^{\dagger}_-V=e^{iE_0t-\Gamma t}\begin{pmatrix}1 &0\end{pmatrix},
\nonumber\\
&u_{\pm}^{\dagger}Vu_{\pm}=0,\qquad u_{-}^{\dagger}Vu_{+}=+ 1,\qquad u_{+}^{\dagger}Vu_{-}=- 1,
\qquad u_{+}u^{\dagger}_{-}V-u_{-}u^{\dagger}_{+}V=I.
\label{6.5}
\end{align}
The appearance of the $-1$ factor in $u_{+}^{\dagger}Vu_{-}$ is not indicative of any possible negative norm ghost problem since  $u_{+}^{\dagger}Vu_{-}$ is a transition matrix element between two different states, and not the overlap of a state with its own conjugate.
As we see, all of the $V$-based inner products are time independent, with the only non-vanishing ones being the ones that link the decaying and growing modes. As the population of one level decreases the population of the other level increases by the same amount, with probability conservation requiring that we consider both levels together and not just one (the decaying one) as is usually done in the  standard Breit-Wigner approach.  

Multiplying the wave functions in (\ref{6.5}) by the spatial terms given in (\ref{5.10a}) then shows that the square well complex conjugate pair probability amplitudes have no exponential growth in either space or time, to thus  recover (\ref{5.13a}).

Given (\ref{6.5}) the associated propagator is  given by \cite{Mannheim2018a,Mannheim2013}
\begin{align}
D_{PT}(E)=\frac{u_{-}^{\dagger}Vu_{+}}{E-(E_0-i\Gamma)}+\frac{u_{+}^{\dagger}Vu_{-}}{E-(E_0+i\Gamma)}=\frac{1}{E-(E_0-i\Gamma)}-\frac{1}{E-(E_0+i\Gamma)}.
\label{6.6}
\end{align}
Thus we obtain
\begin{align}
D_{PT}(E)=\frac{-2i\Gamma}{(E-E_0)^2+\Gamma^2},
\label{6.7}
\end{align}
to thus give the same negative imaginary sign for the propagator as obtained in the standard (\ref{2.2}) that only contained one pole, so that operationally (\ref{6.7}) and (\ref{2.2})  are equivalent, with both giving rise to just one peak in a scattering cross-section at $E=E_0$ with a width $\Gamma$. It is this equivalence that  enabled Lee and Wick \cite{Lee1969} to consider the relativistic generalization of (\ref{6.6}), viz. a propagator of the form $1/(k^2-M^2+iN^2)-1/(k^2-M^2-iN^2)$, in order to obtain better large $k^2$ behavior (viz. $1/k^4$) than a standard $1/k^2$ propagator, to  then control quantum field theory renormalization \cite{footnote6}.

While $D_{PT}(E)$ gives the energy dependence of the propagator,  how we determine its time dependence, viz. 
\begin{align}
D_{PT}(t)=\frac{1}{2\pi }\int^{\infty}_{-\infty} dEe^{-iEt}D_{PT}(E),
\label{6.8}
\end{align}
depends on the form of the  contour that we choose in the complex $E$ plane. There are two poles in $D_{PT}(E)$ as given in (\ref{6.6}), one above the real $E$ axis and one  below, and we can suppress the lower-half circle contribution when $t>0$, and suppress the upper-half circle contribution when $t<0$. However, in order to be able to combine the two propagators in (\ref{6.6}) so as to give the propagator given in (\ref{6.7}), just as in \cite{Lee1969}  we need the two poles in $D_{PT}(E)$ to be in the same contour. We thus deform the contour around the upper-half plane pole at $E=E_0+i\Gamma$ in (\ref{6.6}) so that it contributes when we close below. And with both poles then being in the same contour, via use of (\ref{4.4}) we obtain
\begin{align}
D_{PT}(t)=-i\theta(t)[e^{-iE_0t-\Gamma t}-e^{-iE_0t+\Gamma t}].
\label{6.9}
\end{align}
Thus we obtain both a forward in time growing mode and a forward in time decaying mode, with the time advanced mode not growing backward in time or being acausal, which it would have been had we located the $E=E_0+i\Gamma$ pole in an upper-half plane contour. Thus we obtain a time-advanced, negative-width generalization of the positive-width time delay produced by a potential well. We can associate the time advance  with an energy-dependent phase shift of the form $\tan \delta=-\Gamma/(E_0-E)$, a phase shift equal to $-\pi/2$ at resonance, together with a scattering amplitude and time advance of  the form  
\begin{align}
f(E)\sim e^{i\delta}\sin\delta=- \frac{\Gamma}{E_0+i\Gamma-E},\qquad \Delta T(E)=\hbar\frac{d\delta}{dE}=-\frac{\hbar\Gamma}{[(E-E_0)^2+\Gamma^2]},\qquad \Delta T(E_0)=-\frac{\hbar}{\Gamma}.
\label{6.10}
\end{align}
Thus in $PT$-symmetric quantum mechanics  a time advance (viz. a negative time delay) is natural, with it being equal and opposite to the standard time delay that,  just as discussed in \cite{Mannheim2018a}, must be accompanied by a  corresponding time advance.

\section{Applications to Spontaneous Emission and Elastic Scattering}
\label{S7}

There are two familiar situations in which we can exhibit the implications of the presence of a pair of complex conjugate energies: spontaneous emission from an excited atomic state and elastic scattering. For the atomic case, due to quantum-field-theoretic  electromagnetic fluctuations in the virtual photon cloud that accompanies each atomic electron, atomic energy levels acquire a natural line width $\Gamma$. As discussed in \cite{Mannheim2025}  we must identify the two eigenvalues in (\ref{6.4}) not as those of the ground state and the excited state of an atom    but as the two transition energies, one to excite and populate the upper level and the other to decay and depopulate the upper level, with both transitions being caused by the natural line width that is associated with the virtual photon cloud that accompanies the atomic electron.  With $PT$ symmetry the associated time delay and time advance will cancel each other identically. In the many  $PT$-symmetry experiments that have been performed the emphasis has been on systems with  gain and loss that is balanced, i.e., $E_0+i\Gamma$ and $E_0-i\Gamma$ with the same $\Gamma$. Balancing time advance and time delay is thus in the same vein. In a typical atomic case $E_0$ is order 1 eV and $\hbar/E_0$ is order $10^{-15}$ seconds,  while $\Gamma$ is of order $10^{-6}$ eV, and $\hbar/\Gamma$ is of order $10^{-9}$ seconds. Thus with the time delay and time advance cancelling each other, we find that to within  $ 10^{-15}$ seconds the emission of the photon by the upper level  essentially occurs at the same  instant as the initial absorption of a photon by the lower level with no $10^{-9}$  seconds time delay \cite{footnote8}. Thus even though  the upper level has a width $\Gamma$ (a linewidth that can actually be measured) the lifetime of its decay is not $\hbar/\Gamma$, even though it had always been thought to be so. In the experiments of \cite{Sinclair2022,Angulo2024}, which use ultracold rubidium atoms, an effect confirming this  has been reported, with  the authors of \cite{Sinclair2022} noting that photons are scattered long before the atom’s spontaneous lifetime has elapsed, while the authors of   \cite{Angulo2024}  report detecting a negative time delay \cite{footnoteA}.

In the standard Breit-Wigner discussion of atomic spontaneous emission given in Sec. \ref{S2} consideration was only given to the decay of the excited state with energy $E_0-i\Gamma$. However, the electron-photon interaction that causes the decay is $PT$ symmetric (actually $CPT$ symmetric but the energies involved are too low to create electron-positron pairs, and multiphoton emission is negligible). Thus there must also be an energy $E_0+i\Gamma$ involved in the process. This energy is associated with the excitation of the ground state into the first excited due to the absorption of an incoming photon. I.e., due to the absorption of the incoming photon into the virtual photon cloud that accompanies the ground state electron, the same virtual photon cloud from which the spontaneous photon emission from the  excited atomic state occurs. We thus identify an apparently previously unrecognized role for $PT$ symmetry in atomic transitions.

In the analogous $\pi^++p\rightarrow \pi^++p$ elastic scattering process the lower level is a free $(\pi^+,p)$ system and the upper level is the $\Delta^{++} (1232)$ resonance with a 120 MeV width. The excitation of the $\Delta^{++}$ is associated with $E=E_0+i\Gamma$ and its decay  with $E=E_0-i\Gamma$, with compensating time advance and time delay. Despite there being two poles in (\ref{6.6}) there is only one upper level and thus just one observable $\Delta^{++}$ (c.f. (\ref{6.7})) and not two.  In regard to  resonances in scattering processes we note also that in searches for where in energy they might be located one does a partial wave analysis and looks for phase shifts that go through $\delta=\pi/2$. This type of search is not affected by the existence of phase shifts that go through $\delta =-\pi/2$, as such a search would only reproduce the same set of resonances as in the $\delta=\pi/2$ case.

In both of these examples the existence of complex energies is due to the presence of a continuum of non-normalizable states. For the $\pi^++p$ elastic  scattering experiment the non-normalizable states are the incoming and outgoing scattering states themselves. However, for the atomic example the non-normalizable continuum states are the virtual  momentum states in the virtual photon cloud that accompanies each electron, with the incoming photon being absorbed by the cloud and the outgoing photon being emitted spontaneously by the cloud that accompanies the excited atomic state. We note that even though the QED fine-structure constant is real, the natural line width arises because of the interaction  of the electron with an infinite set of virtual photon momentum-eigenstate plane waves. So just like the square well, having a real coupling constant is not sufficient to prevent the presence of  non-stationary states.

Thus to summarize we note that in the cases described here and in cases like them probability is conserved by the interplay between the growing and decaying modes, with the only non-vanishing matrix elements in (\ref{6.5}) being precisely of this time-independent form.

\section{Implications for Scattering and $PT$ Symmetry}
\label{S8}

Our analysis provides an extension of the way that an antilinear symmetry such as $PT$ symmetry can be  realized. The two ways that $PT$ symmetry has been realized in the literature are for systems with a finite number of degrees of freedom and  systems with an infinite number of degrees of freedom.  The finite-dimensional situation can be described by the illustrative matrix 
\begin{align}
M(s)=\begin{pmatrix}
1+i&s\\ 
s&1-i
\end{pmatrix},
\label{8.1}
\end{align}
where the parameter $s$ is real and positive. The matrix $M(s)$ does not obey the Hermiticity condition $M_{ij}=M^*_{ji}$. However, if we set $P=\sigma_1$ and  $T=K$ where $K$ denotes complex conjugation we obtain $PTM(s)[PT]^{-1}=M(s)$, with $M(s)$ thus being  $PT$ symmetric for any value of  the real parameter $s$. With the eigenvalues of $M(s)$ being given by $E_{\pm}=1 \pm (s^2-1)^{1/2}$, we see that both of these eigenvalues are real if $s$ is either greater or equal to one, and form a complex conjugate pair if $s$ is less than one, just as is to be expected with $PT$ symmetry. In all of these realizations the matrix remains two-dimensional, and can be taken to act on one and the same Hilbert space that is spanned by  spin up (transpose of $(1,0)$) and spin down (transpose of $(0,1)$) vectors. The matrix  $M(s>1)$ can be brought to a Hermitian form by a similarity transformation \cite{footnote10}, and the  $M(s<1)$ matrix can be brought to a diagonal form (viz. (\ref{6.4}) with $E_0=1$, $\Gamma=(1-s^2)^{1/2}$) by a similarity transformation \cite{footnote11}. 

However, when $s=1$ the $M(s=1)$ matrix only has one right-eigenvector, viz. the transpose of $(1,-i)$ with eigenvalue $1$, and one left-eigenvector, viz. $(i,1)$ with eigenvalue 1. In the Schr\"odinger equation $id\psi/dt=M(s=1)\psi$ the right-eigenvector is the transpose of $(e^{-it},-ie^{-it})$, and in $-id\psi/dt=\psi M(s=1)$ the left-eigenvector is $(ie^{it},e^{it})$. The other two states, both not eigenvectors of $i\partial_t$ are a right-vector that is the transpose of $((1+t)e^{-it},-ite^{-it})$, and a left-vector of the form $(ite^{it},(1+t)e^{it})$, solutions  that, just like $\psi_2(r,t)$ of (\ref{5.16a}),  grow linearly in time. Just as in (\ref{6.5}) all overlaps are time independent, as has to be the case since with $V=\sigma_1$, $\sigma_1M(s)\sigma_1=M^{\dagger}(s)$, (\ref{6.2}) is satisfied.
Thus to conclude we see that the general $M(s)$ example with $s>1$, $s=1$ and $s<1$ sectors shows that $PT$ symmetry includes Hermiticity while also generalizing it.

For the square well the $s>1$ case corresponds to the bound state sector with real eigenvalues,  the $s<1$ case corresponds to the above threshold scattering sector with complex conjugate energy eigenvalues, while  the $s=1$ case corresponds to the scattering at threshold with the branch point being an exceptional point at which there are  states that are not eigenstates of the Hamiltonian, states that belong to a basis in which the Hamiltonian  cannot be diagonalized. 

For the infinite degree of freedom situation a typical example is the Pais-Uhlenbeck (${\rm PU}$) two-oscillator model. The model is the non-relativistic limit of a covariant fourth-order derivative neutral scalar field theory with the linear momentum frozen out  \cite{Mannheim2018a}, and has action and  Hamiltonian of the form \cite{Mannheim2000,Bender2008a}  
\begin{align}
I_{\rm PU}&=\frac{1}{2}\int dt\left[{\ddot z}^2-\left(\omega_1^2
+\omega_2^2\right){\dot z}^2+\omega_1^2\omega_2^2z^2\right],
\nonumber\\
H_{\rm PU}&=\frac{p^2_z}{2}+p_zx+\frac{1}{2}\left(\omega_1^2+\omega_2^2 \right)x^2-\frac{1}{2}\omega_1^2\omega_2^2z^2,
\label{8.2}
\end{align}
where $[z,p_z]=i$ and  $[x,p_x]=i$. If we set up a Schr\"odinger problem by setting $p_z=-i\partial/\partial_z$, $p_x=-i\partial/\partial_x$ we find that the state with energy $\omega_1+\omega_2$ has an eigenfunction of the form 
\begin{align}
\psi(z,x)&=\exp\bigg[\frac{1}{2}(\omega_1+\omega_2)\omega_1\omega_2z^2
+i\omega_1\omega_2zx-\frac{1}{2}(\omega_1+\omega_2)x^2\bigg].
\label{8.3}
\end{align}
With $\omega_1+\omega_2$ and $\omega_1\omega_2$ taken to be real and positive this eigenfunction is normalizable in $x$ but not in $z$. We therefore continue the coordinate $z$ into the complex plane into two so-called Stokes wedges that are contained in the north and south quadrants of a letter $X$ that is drawn in the complex $z$ plane. In these two wedges $\psi(z,x)$ is square integrable. To implement the continuation for operators we make similarity  transformations of the form $e^{\pi p_zz/2}ze^{-\pi p_zz/2}=-iz:=-iy$,
$e^{\pi p_zz/2}p_ze^{-\pi p_zz/2}=ip_z:=iq$ with $[y,q]=i$. In terms of $y$
and $q$ the Hamiltonian and $\psi(z,x)$ take the form
\begin{align}
e^{\pi p_zz/2}H_{\rm PU}e^{-\pi p_zz/2}&=H^{\prime}_{\rm PU}=\frac{p_x^2}{2}-iqx+\frac{1}{2}\left(\omega_1^2+\omega_2^2
\right)x^2+\frac{1}{2}\omega_1^2\omega_2^2y^2,
\nonumber\\
\psi(y,x)&=\exp\bigg[-\frac{1}{2}(\omega_1+\omega_2)\omega_1\omega_2y^2
-\omega_1\omega_2yx-\frac{1}{2}(\omega_1+\omega_2)x^2\bigg],
\label{8.4}
\end{align}
with the other eigenfunctions being polynomial functions of $y$ and $x$ times $\psi(y,x)$.
As constructed, because of the $-iqx$ term $H^{\prime}_{\rm PU}$ is not Hermitian. However,  if we assign $PxP^{-1}=-x$, $Pp_xP^{-1}=-p_x$, $PyP^{-1}=y$, $PqP^{-1}=q$, $TxT^{-1}=x$, $Tp_xT^{-1}=-p_x$, $TyT^{-1}=-y$, $TqT^{-1}=q$, we find that $H^{\prime}_{\rm PU}$ is $PT$ symmetric. 

For $H^{\prime}_{\rm PU}$ there are three realizations of interest: (i) $\omega_1$ and $\omega_2$ both real, positive and unequal, (ii)  $\omega_1$ and $\omega_2$ both real, positive  and equal, and (iii) $\omega_1=a+ib$ and $\omega_2=a-ib$, where $a$ and $b$ are both real and positive. These three realizations respectively correspond to (i) all energies real (just the infinite set of excitations of a two-dimensional harmonic oscillator) \cite{Bender2008a}, (ii) a non-diagonalizable case with both stationary and non-stationary solutions \cite{Bender2008b}, and (iii) one no-particle state with energy $2a$ and the rest either real or in complex conjugate pairs of eigenenergies \cite{Mannheim2018a}. The three of these realizations provide just what is to be expected of a $PT$-symmetric Hamiltonian. In all three of these realizations $\omega_1+\omega_2$ and $\omega_1\omega_2$ are real and positive, with all of the $H^{\prime}_{\rm PU}$ eigenfunctions thus being square integrable in all of them. Thus all three cases are well defined in the self-same Stokes wedges, so that as we continue the frequencies between the various cases we always remain in the same Hilbert space.

In passing we note that an interesting aspect of the ${\rm PU}$ complex conjugate pair realization with $\omega_1=a+ib$ and $\omega_2=a-ib$ is that in it the initial $H_{\rm PU}$ Hamiltonian  as given in (\ref{8.2}) takes the form 
\begin{align}
H_{\rm PU}=\frac{1}{2}p^2_z+p_zx+(a^2-b^2)x^2-\frac{1}{2}(a^2+b^2)^2z^2.
\label{8.5}
\end{align}
Thus despite the presence of complex energies all of the coefficients in the Hamiltonian are real, just like the square well, so we do not have to have a complex potential in order to obtain complex energies. There is not only no need for any complex potential in the complex energy realization of the ${\rm PU}$ two-oscillator model, there is also no need for any dissipative terms such as those that are used for the damped harmonic oscillator, with there being no term in the $I_{\rm PU}$ action given in (\ref{8.2}) that is linear in the velocity.  Also we note that even with complex potentials there can still be real solutions, with all the solutions to $\hat{H}=\hat{p}^2+i\hat{x}^3$ being real \cite{Bender1998}, though  as shown in \cite{Bender1998} $\hat{H}=\hat{p}^2+i\hat{x}$ has complex conjugate pair solutions \cite{footnote12}.

Unlike the above examples, the behavior of a system with bound states with real $E$ below the threshold energy $E_{\rm THR}$ and scattering states when ${\rm Re}[E]>E_{\rm THR}$ is quite different. For the bound states the wave functions are square integrable while the scattering state wave functions are not. The bound state wave functions are in a square-integrable Hilbert space while the scattering state wave functions are not.  In the complex $E$ plane there is a branch point  at $E_{\rm THR}$. In the continuation from $E<E_{\rm THR}$ to  $E>E_{\rm THR}$, at the branch point the wave functions leave the Hilbert space of square-integrable wave functions.  In addition, we note that while $\delta (p)$ involves a continuation of coordinates into the complex plane (as per (\ref{4.1}) and its $\delta(p)$  derivative) this is not the same as a continuation of a wave function into an appropriate Stokes wedge,  with there being no Stokes wedge domain in the complex coordinate plane in which a plane wave is square integrable. We thus enlarge the way that an antilinear symmetry such as $PT$ symmetry can be  realized.

\section{Final Comments}
\label{S9}
 When a process with a  Hamiltonian that is thought to be Hermitian has energy eigenvalues that are not real,  it is generally thought that the system being considered is only a subsystem of some larger system (with the subsystem being referred to as being open). And that when the larger system is also included Hermiticity and probability conservation will be restored.  That is not what is being considered in this paper. Rather, the presence of complex energies in the scattering sector is because of a breakdown of the relation between Hermiticity and real eigenvalues, as that relation does not hold for states that grow exponentially in space, and that is what allows there to be complex energies. Thus while a Hamiltonian might be self-adjoint in the bound state sector it is not self-adjoint in the scattering sector, just as we find with the square well. Nonetheless, it can still have an antilinear symmetry, and in fact, because of the generality of the $CPT$ theorem,  the Hamiltonian must have an antilinear symmetry in both sectors. Thus we find that the square well admits of energy eigenstates with eigenvalues $E_0\pm i\Gamma$ that are in complex conjugate pairs, just as antilinear symmetry requires. This translates into $E_0\pm i\Gamma$ poles in the scattering amplitude, to thereby give a structure to the scattering amplitude that had not previously been considered. Thus Hermiticity is not an intrinsic property  of a Hamiltonian but a basis-dependent one, as it is determined by whether or not it is self-adjoint in any given basis. However antilinear symmetry of a Hamiltonian is an intrinsic property of an operator, regardless of any self-adjointness issues. Thus as we proposed in \cite{Mannheim2018a} antilinearity is a more general guideline for quantum theory  than Hermiticity.

 With such an antilinear symmetry there can be exceptional points in  which there are states that are not eigenstates of the  Hamiltonian. By explicit calculation we have shown the spherically symmetric three-dimensional square well potential possesses such exceptional points for those values of $V_0$ that obey $V_0=\pi^2\hbar^2/8ma^2,~ 9\pi^2\hbar^2/8ma^2,~25\pi^2\hbar^2/8ma^2,\ldots$ In these particular cases the threshold branch point is an exceptional point. 

While there can be complex energies in the scattering sector, in and of itself that is not sufficient to ensure probability conservation as states with energies of the form $E_0-i\Gamma$ decay and take  probability with them. If such systems are open systems the inclusion of the larger system in which they would then be contained could restore probability conservation (or Hermiticity).  However, with antilinear symmetry energies of the form $E_0+i\Gamma$ are contained within the system that contains the $E_0-i\Gamma$ modes. Systems that contain both the $E_0-i\Gamma$ and $E_0+i\Gamma$ modes are not open but are  self-contained closed systems. The loss of population of a level as it decays ($E_0-i\Gamma$) leads to growth ($E_0+i\Gamma$) in the population of the level that it decays into. Probability conservation then requires that the total population of the two levels combined be constant. While radioactive decay precedes growth in the population of  the decay products  in the Rutherford case, probability conservation can be maintained if the growth occurs first, this being the case in both spontaneous atomic emission and elastic scattering.

The $\pi^++p$ elastic scattering system that we discussed above is an example of a closed not an open system since below the inelastic threshold all that is involved is the elastic $\pi^++p$ channel. The scattering produces a $\Delta^{++}$ resonance at 1232 MeV with width 120 MeV as the $\pi^+$ and $p$ come together to excite the $\Delta^{++}$ (viz. $E_0+i\Gamma$) and then separate as the $\Delta^{++}$ decays ($E_0-i\Gamma$). The scattering process thus involves not one but two poles in the scattering amplitude (cf. (\ref{6.6})), one for excitation of the $\Delta^{++}$ and one for its decay. Both poles are needed for probability conservation, with there being only one observable $\Delta^{++}$ (cf. (\ref{6.7})) and  not two. The description of resonance production that is presented here does not appear to have previously been considered in the literature.

In both  elastic scattering and scattering off a square well  a continuum of momentum eigenstates is automatically present since the plane wave modes are on shell. However, what makes the spontaneous atomic emission case discussed above so interesting  and novel is that there the needed continuum is virtual, residing within the photon cloud that accompanies an electron.

Our study of the finite-depth square well has revealed that modes with energy $E_0-i\Gamma$ have a spatial wave function that grows exponentially (complex energy entails complex momentum), to thus be Gamow vectors. And if, as had been widely thought,  they were to be the only above-threshold poles in the scattering amplitude the theory would have had to be rejected unless the Hilbert space is appropriately rigged.  However, because of $PT$ symmetry there are companion states with energy $E_0+i\Gamma$, and they have a spatial wave function that falls exponentially (anti-Gamow vectors). Then again because of $PT$ symmetry one has to use the $\hat{V}$ norm where $\hat{V}$ is the operator that effects $\hat{V}\hat{H}\hat{V}^{-1}=\hat{H}^{\dagger}$, and this norm has no exponential growth in either space or time. Thus  because of $PT$ symmetry the square well can be made viable without rigging, with the anti-Gamow vector serving as what would otherwise be the test function in a rigged Hilbert space. As a model the square well has served as a prototype for scattering by attractive short-range potentials. Amending it as done here through $PT$ symmetry enables the square well to continue to do so. Finally, to conclude, we note that any system for which the finite-depth square well is a prototype (going back to as early as the first discussions of the finite-depth square well itself in quantum mechanics but excluding systems with an infinitely high potential barrier) is actually an antilinear theory rather than a Hermitian theory, with $PT$-type theories thus having existed long before they were identified as such.
\medskip

\noindent
Acknowledgment: 
The author wishes to acknowledge helpful discussions with Dr. C. M. Bender.

\bigskip
\noindent
Data Availability Statement:  Data sharing is not applicable to this article as the only data used are presented in \cite{footnote00},


\begin{thebibliography}{99}
\bibitem{footnote0} Some of the analysis  that we present here duplicates that of \cite{Mannheim2018a} and the short letter in \cite{Mannheim2025} and is included here as it provides the background needed for this paper, and enables this paper to be self-contained and complete.
\bibitem{footnote00} Setting $\gamma=1$, $a=1$ and $V_0=16$, from (\ref{5.7a}) we obtain $\mu=7.661$, $\nu=1.288$, $E_0=57.025$, $\Gamma=19.738$, $K=7.661\pm 1.288i$, $k=6.579\pm 1.500 i$. Inserting the real parts of $K$ and $k$ into (\ref{5.2}) gives $\tan\delta =1.748$, which is far from  the infinite $\tan(\pi/2)$. From (\ref{5.2}) the real $K$ and $k$ that would give $\tan\delta =\infty$, i.e., give $\beta\tan\alpha=-1$, are $K=6.045$, $k=4.532$, with $E_0=36.542$. With $\beta=-1/\tan\alpha$,  at $k=4.532$ we obtain $\beta_0=-0.182$, and from (\ref{5.21a}) we obtain the negative $\Gamma=-5.666$. The $\tan\delta=-i$ $PT$-theory  $E_0=57.025$, $\Gamma=19.738$ pair is thus very different from the $\tan\delta=\infty$ Breit-Wigner $E_0=36.542$, $\Gamma=-5.666$ pair.
\bibitem{footnote8} Given the presence of the  causal $\theta(t)$ factor in (\ref{6.9}) an excited state does not emit a photon before the ground state absorbs it, i.e., the time advance is not bigger than the time delay.
\bibitem{footnoteA} The $PT$-symmetry time advance discussed here is not the same as a familiar time advance effect  that  occurs in quantum mechanics. In that effect  the center of a wave packet approaching  a well (or barrier) travels on the classical trajectory. However, the wings of the wave packet not only spread out they can start to deform due the presence of the well before the center of the wave packet actually reaches the well.
\bibitem{footnote11} For  the complex conjugate eigenvalue case with $s<1$ we can diagonalize $M(s<1)$ via $SM(s<1)S^{-1}$ where $S=\cosh\alpha+\sigma_2 \sinh\alpha$ with $\tanh(2\alpha)=s$, and obtain (\ref{6.4}) as written with $E_0=1$ and $\Gamma=(1-s^2)^{1/2}$.
\bibitem{footnote12}  It was through study of systems such as these that  a role for  $PT$ symmetry in quantum theory was identified, with any theory with an antilinear symmetry then generically being referred to as a $PT$ theory.
\end{thebibliography}
\end{document}